\documentclass[letterpaper,12pt]{article}

\usepackage{amsfonts}
\usepackage{amstext}
\usepackage{epsf}
\def\W#1{\ ^{W}\!\!#1}

\title{Understanding Quantum Theory in Terms of Geometry}
\author{Fatimah Shojai$^1$\footnote{fatimah@ipm.ir}\\
and\\ Ali Shojai$^{1,2}$\footnote{shojai@ipm.ir}\\
$^1$ Institute for Studies in Theoretical Physics and Mathematics,\\
POBox 19395--5531, Tehran, Iran.\\ $^2$ Physics Department, Tehran University,\\
Tehran, Iran.\\ To appear in \textit{Progress in Quantum Physics Research, Nova Science Publishers, Inc., 2004}}
\date{}
\begin{document}
\maketitle
\tableofcontents
\begin{abstract}
Understanding quantum theory in terms of a geometric picture sounds great. There are different approaches to this idea. Here we shall present a geometric picture of quantum theory using the de-Broglie--Bohm causal interpretation of quantum mechanics. We shall show that it is possible to understand the key character of de-Broglie--Bohm theory, the quantum potential, as the conformal degree of freedom of the space--time metric. In this way, gravity should give the causal structure of the space--time, while quantum phenomena determines the scale. Some toy models in terms of tensor and scalar--tensor theories will be presented. Then a few essential physical aspects of the idea including the effect on the black holes, the initial Big--Bang singularity and non locality are investigated. We shall formulate a quantum equivalence principle according to which gravitational effects can be removed by going to a freely falling frame while quantum effects can be eliminated by choosing an appropriate scale. And we shall see that the best framework for both quantum and gravity is Weyl geometry. Then we shall show how one can get the de-Broglie--Bohm quantum theory out of a Weyl covariant theory. Extension to the case of many particle systems  and spinning particles is discussed at the end.
\end{abstract}
\section{Introduction and survey}
In this century, physicists have been departed from 19th century physics, in two ways. The first was the generalization and bringing the old idea of \textit{frame independence} or \textit{general covariance}, in a manifest form. The result of this effort was the pioneer general relativity theory, in which the gravitational effects
of matter are identified with the geometry of the space--time.  The enigmatic character of this theory is just the above-mentioned property, i.e. the interconnection of gravity and general covariance. When one tries to make a general covariant theory, one is
forced to include gravity.

The main root of this interconnection is the \textit{equivalence principle}. According to the equivalance principle, it is possible to go to a frame in which gravity is locally
absent, and thus the special theory of relativity is applicable locally. Now using the general covariance and writing down anything in a general frame, we will get the general relativity theory\cite{weinberg}.

The second was the investigation of the quantal behavior of matter, that leads to the \textit{quantum theory},  according to which a great revolution  appeared in physics. In order to explain the atomic world, the quantum theory threw out two essensial classical concepts, the \textit{principle of causality} and \textit{the dogma of formulation of physics in terms of motion in  space--time} (motion dogma). The
first one is violated during a measurement process, while the second does not exist at any time.

After the appearance of quantum mechanics, it was proven that not only do the ordinary particles show quantal behavior but mediators of the fundamental forces also do so. In this way quantum electrodynamics, quantum chromodynamics and quantum flavor dynamics were born. But the construction of quantum gravitodynamics or quantum gravity, and its application to cosmology, is considerably very problematic\cite{ash}. These difficulties may be mainly divided into two categories. Some of them are related to the conceptual problems of the standard quantum mechanics, while others are specific to gravity, and are in fact related to the classical features of gravity theory. The first category includes the measurement problem and the meaning of the wave function of the universe, while the vanishing of the hamiltonian which leads to the time independence of the wave function and nonrenormalizability, belong to the second category.

From a fundamental physical viewpoint, in contrast to the general theory of relativity which is the best theory for gravity, standard quantum mechanics is not the only satisfactory way of understanding the quantal behavior of matter.  One of the best theories explaining the quantal behavior of matter but remaining faithful to the principle of causality and the motion dogma, is the de Broglie--Bohm quantum theory.\cite{bohm} According to this theory, all the enigmatic quantal behavior of the matter results from a self-interaction of the particle. In fact, any particle which exerts
a \textit{quantum force} on itself can be expressed in terms of a \textit{quantum potential} and which is derived from the particle wave function.

The celebrated property of the de Broglie--Bohm quantum theory is the following property. \textit{At anytime, even when a measurement is done, the particle is on the trajectory given by Newton's law of motion, including the quantum force}. During a measurement, the system is in fact a many-body system (including the particle itself, the probe particle, and the registrating system particles). When one writes down the appropriate equation of motion of all the particles,\footnote{There is a consistent de Broglie--Bohm quantum theory for a many particle system} and when one considers
the very fact that we know nothing about the initial conditions of the registerating system particles, one sees how the projection postulate of quantum mechanics came about\cite{bohm}. Accordingly the result of any measurement is one of the eigenvalues of the operator related to the measured quantity with some calculable probability
distribution. The de-Broglie--Bohm quantum theory of motion, is a causal theory which although behaves as the Copenhagen quantum mechanics at the statistical level, it has non of the conceptual difficulties of the standard quantum mechanics. It is well proved that the causal theory reproduces all the results of the orthodox quantum theory\cite{bohm}, as well as predicting some new results (such as time of tunneling through a barrier\cite{cush}) which in principle lets the experiment to
choose between the orthodox and the causal quantum theories. Perhaps the most important point about the causal theory is that it presents a causal deterministic description of the reality. So it looks very natural to make a quantum theory of gravity in the spirit of the de-Broglie--Bohm quantum theory of motion. In the standard form of this theory, the classical gravity should be viewed as a field. Then it is possible to construct Bohmian metric trajectories . This is what is essentially done in \cite{gil}.

A way of struggling with quantum gravity is to use the \textit{minisuperspace} of the conformal degree of freedom of the space-time metric\cite{nar}. This approach has several fruitfull results. It admits non-perturbative calculations, and it is very useful for studying quantum cosmology. Because the isotropic and homogeneous space-time used in   cosmology is conformally flat. In addition, by including the effects of the back--reactions of the quantum variable (i.e. the conformal factor) on the background metric, one arrives at some extended form of Einstein's equations. These semi-classical equations lead to non--singular cosmological solutions and they have the correct classical limit. In this approach, by merely quantizing one degree of freedom of the space-time metric, and by considering the back--reaction effects, the time independence problem is solved. This is achieved because of the extension of Einstein's equations. But it must be noted that the physical meaning of the quantum variable, i.e. the conformal factor, is not clear in this approach. The non-singularity
of the results of the above approach rests on the theorem\cite{kem} which states that \textit{for any singular metric , there is some appropriate conformal factor, in such a way that conformal metric is non-singular}.

The present work tries to combine the de-Broglie--Bohm quantum theory of motion
and gravity in a very different way. The foundation of this approach is the de-Broglie
remark\cite{deb} that \textit{the quantum theory of motion for relativistic spinless particles is very similar to the classical theory of motion in a conformally flat space-time}. The conformal factor is related to the Bohm's quantum potential. We shall present a generalization and an appropriate formulation of this remark. That is to say, we geometrize Bohmian mechanics according to the de-Broglie remark. Then, it can be seen that the effects of gravity on geometry and the quantum effects on the geometry of the space-time are highly coupled. In fact there are two contributions to the background metric the gravitational quantal effects of matter which constitute the energy--momentum tensor. Since in the evaluation of the quantal part the background metric is used, the gravitational and quantal contributions to the background metric are so highly coupled that no one without the other has any physical significance.

It must be pointed out here that as aresult the conformal factor is meaningless as the enesemble density goes to zero and the geometry looses its meaning at this limit. This is a desired property, because it is in accord with Mach's principle, which states that for an empty universe the space--time should be meaningless. In subsection(\ref{s1})\cite{geo} the authors, as a first step towards the formulation of the above conclusion, introduced the quantum conformal degree of freedom via the method of Lagrange multipliers. In this way there are a set of equations of motion describing the background metric, the
conformal degree of freedom and the particle trajectory. A corollary of this theory is that one can always work in a gauge (classic gauge) in which no quantum effect be present or in a gauge (quantum gauge) in which the conformal degree of freedom of
the space--time metric is identified with the quantum effect This, in its turn, leads to dramatic departures from the classical prediction, when both the effects of gravity and quantum on geometry are considerable, i.e. around those areas of the space-time which are singular according to the classical theory.

As a different approach in ref \cite{mot} the authors symmetrized the Brans--Dicke theory by a conformal transformation. And arrive to a particle interpretation suggesting that the quantum aspect of matter can be geometrized.

In \cite{conf}, the conformal transformation was applied only  to the space--time metric. Other quantities like mass, density and so on were assumed to posses no
 transformation. This is because the above conformal transformation which incorporates the quantum effects of matter into a specific conformal factor, is in fact a scale transformation. As the conformal transformation is more general than scale  transformation which is used in \cite{geo}, it seems preferable  to make a conformal transformation, in  which all physical quantities are transformed, instead of making only a scale transformation.
In reference \cite{conf}, it is shown that by the conformal transformation the equation of motion would be transformed to an equation in which there is no quantum
effects. As a result, the geodesic equation would be changed to the one without the quantum force. This means that it is possible to have two identical pictures for investigating the quantal effects of matter in the curved space--time background. According to the first picture, the space--time metric contains only the gravitational effects of matter. The quantum effects affect the path of the particles via the quantum force. In the second picture, the space--time metric is related to the previous by a conformal factor and contains the gravitational and  quantal effects of matter.

This shows that the quantum as well as the gravitational effects of matter have geometrical nature. The second picture mentioned above provides a unified geometrical framework for understanding the gravitational and quantum forces. Accordingly, we call the conformal metric as the physical metric (containing both gravity and quantum) and the other metric is the background metric (including only gravity).

The above-mentioned theory,\cite{geo} has a problem.  In this theory, it is assumed that one deals with an ensemble of similar particles with density. In Bohm's theory, the quantum potential exists for a single particle as well as for an ensemble. In the case of a single particle, the interpretation of the quantum
potential is in terms of an hypothetical ensemble.  Note that in the above theories, the ensemble is a real one, not an hypothetical one, because, the energy--momentum tensor of the ensemble is appeared and has physical effects. As we shall show in subsection (\ref{s2})\cite{qgg}, we have solved this problem and the theory would work both for a single
particle and for an ensemble.

In subsection (\ref{s2}) we shall show that it is possible to make a pure tensor theory for quantum gravity. As a result we shall show that the correct quantum conformal degree of freedom would be achieved, and that the theory works for a particle as well as for a real ensemble of the particle under consideration and that it includes the pure quantum gravity effects. We shall do all of these by trying to write the quantum potential terms in terms of geometrical parameters, not in terms of ensemble properties.

The important point about both references\cite{geo} and \cite{conf} is that in order to fix the relation of the conformal degree of freedom of the space--time metric and the quantum potential, the method of lagrange multiplier is used and in this way they are a
little artificial. In subsection (\ref{s3}) we shall show\cite{st} that in the framework of the scalar--tensor theories, it is possible to write an action principle, in which both
gravitational and quantum contributions to the geometry are included and that the
conformal degree of freedom of the space--time metric is fixed at the level of the
equations of motion not needing the method of lagrange multiplier.

Next in subsection (\ref{s4})\cite{ss} we attend to the double scalar case because in some theories such as superstring and Kaluza--Klein, it is more useful\cite{bar}. In both of these  theories the gravitational interaction includes two other fields in addition to the metric field. In string frame(or Jordan frame)one of them is coupled nonminimally to gravity
as in the Brans--Dicke theory and the other is coupled minimally to gravity, but has a nontrivial coupling with the first scalar field. Note that, in these theories one can couple both the scalar fields minimally to gravity by a conformal transformation (Einstein frame). As a result, the question that the physical interpretation must be presented in which frame, is an open problem\cite{far}. On the other hand we shall show that using two scalar fields, one can relax this preassumption and on the equations of motion, the correct form of quantum potential will be achieved.

In subsections (\ref{s5}), (\ref{s6}) and (\ref{s7}) some general solutions are obtained. using these, the important question that if this quantum gravity theory leads to some new results, is investigated. That solusions are used for black holes and bigbang in subsections (\ref{s8}) and (\ref{s9})\cite{qgg}. In subsection (\ref{s10})\cite{clu} we are interested in investigating whether this theory has anything to do with the cluster formation or clustering of the initial uniform distribution of matter in the universe. The problem of cluster formation is an important problem of cosmology and there are several ways to tackle with it\cite{bor}. Here we don't want to
discuss those theories, and our claim is not that the present work is a good one. Here we only state that \textit{the cluster formation can also be understood in this way}. It is a further task to decide if this work is in complete agreement with experiment or not.

A special aspect of the quantum force is that it is highly nonlocal. This property, is an experimental matter of fact \cite{bell}. Since the mass field represents the conformal degree of freedom of the physical metric, quantum gravity is expected to be highly nonlocal. In the subsection (\ref{s11})\cite{non} this is shown explicitley for a specific problem.

From a different point of view it has been believed for a long time that the long range forces (i.e. electromagnetism and gravity) are different aspects of a unique phenomena. So they must be unified.
Usually it is proposed that one must generalize Einstein's general relativity theory to have a geometrical description of electromagnetic fields. This means to change the properties of the manifold of general relativity. Using higher dimensional manifolds\cite{wes}, changing the compatibility relation between the metric and the affine connection\cite{we} and using a non-symmetric metric\cite{ein} are some examples of the attempts towards this idea. In all the above approaches, the additional degrees of freedom correspond  to the components of the electromagnetic potential.
The second idea leads to the Weyl's gauge invariant geometry. Apart from the electromagnetic aspects of Weyl geometry, it has some other applications. Some authors believe that Weyl geometry is a suitable framework for quantum gravity.
E.g. in a series of papers \cite{odin} a succesful approach to Weyl quantum gravity and conformal sector in quantum gravity is presented. The authors have used an effective theory based on integrated conformal anomaly dynamics, in the infrared region. They also have considered a sigma model action which is the most general version of a renormalizable theory in four dimmensions. They have investigated the phase structure and the infrared properties of conformal
quantum gravity and then extend its results to higher derivative quantum gravity.

Also in ref\cite{whe} a new quantum theory is proposed on the basis of Weyl picture which is purely geometric. The observables are introduced as zero Weyl weight quantities. Moreover any weightful field has a Weyl conjugate such as complex
conjugate of the state vector in quantum mechanics. By these dual fields, the probability can be defined. These are the elements of a consistent quantum theory which is equivalent to the standard quantum mechanics. Moreover it is shown that the quantum measurement and the related uncertainty  would emerged from Weyl
geometry naturally. In this theory when the curl of Weyl vector is zero, we arrive at the classical limit. By noting the transformation relation of Weyl vector, it is concluded that the change of length scale is only a quantum effect.

One more approach to geometrize quantum mechanics can be found in \cite{wood}. Here a modified Weyl--Dirac theory is used to join the particle aspects of matter and Weyl symmetry breaking. This is also a geometrization of quantum mechanics. Also one can find the relation of quantum potential, the basic character of Bohm's theory, to the fundamental geometric properties, especially to the curvature of the space-time using Weyl geometry in \cite{san}. Furthermore in \cite{sid} Sidharth considers the geometrical interpretation of quantum mechanics from the point of view of non-commutative non-integrable geometry.

In the present work we shall look at the conformal invariance at the quantum level. Does the quantum theory lead us to any characteristic length scale and thus break the conformal symmetry? Or conversely the quantum effects lead us to a conformal invariant geometry? In section (\ref{s12}) we shall discuss these questions in the context of the causal quantum theory proposed by Bohm\cite{bohm} and use our new way of
geometrization of quantum mechanics introduced in here. We emphasize that what we shall show that our specific geomerization of quantum mechanics procedure (based on Bohmian quantum mechanics) can be better understood in the Weyl framework. This is different from Weyl quantum gravity approaches like those of \cite{odin}.

We shall show that the Weyl vector and the quantum effects of matter are connected. We shall see how the
conformal symmetry emerges naturally by considering quantum effects of matter. Finally in section (\ref{s13}) we show that the Weyl--Dirac theory is a suitable framwork for
identification of  the conformal degree of freedom of the space--time with the Bohm's quantum mass.

From a similar perspective, Quiros and et all\cite{quir} discuss the space--time singularity by the geometrical dual representation in general relativity. On this basis they emphasis on the Weyl integrable geometry as a consistent framework to describe the gravitational field.

Finally in section (\ref{s20}) we shall investigate possible extension of our results in two ways. First analyzing the case of many--particle systems and second, inclusion of spin.
\section{The geometric nature of quantum potential}
\subsection{Non--relativistic de-Broglie--Bohm theory}
The de-Broglie--Bohm quantum theory of motion\cite{bohm} is a causal theory which
although agrees with the Copenhagen quantum mechanics at the statistical level,
it is able to determine the exact path of a particle. In this way it predicts
all the physical quantities of a particle, \textit{deterministically}.
This theory does not contain such difficulties as the reduction of the wave function,
and so on \cite{bohm}. Therefore, it seems more appropriate that in building  a quantum
theory in the presense of gravity, to use the de-Broglie--Bohm theory rather than the Copenhagen
quantum mechanics. Because in this case some of the conceptual
problems of the standard
quantum mechanics appear more clearly\cite{haw}.

Now, we make a glance at the de-Broglie--Bohm
quantum theory of motion. It contains three postulates\cite{bohm}.
The first one states that for any particle, there is an \textit{objectively real} field
$\psi(\vec{x},t)$ which in the non-relativistic domain satisfies the Schr\"odinger equation.
The second postulate presents the effect of the field on the particle. According
to this postulate the linear momentum of the particle is given by the so--called
\textit{guidance formula}:
\begin{equation}
\vec{p}=\vec{\nabla}(\hbar \times \text{phase of}\  \psi)
\label{cc}
\end{equation}
One can show that the particle expriences the force
\begin{equation}
\vec{F}=-\vec{\nabla}Q
\label{aa}
\end{equation}
from the field $\psi$, where $Q$ is the \textit{quantum potential} given by:
\begin{equation}
Q=-\frac{\hbar^2}{2m}\frac{\nabla ^2 |\psi|}{|\psi|}
\end{equation}
Finally the third postulate states that at the statistical level we have:
\begin{equation}
\rho=\psi^*\psi=|\psi|^2
\end{equation}
where $\rho$ is the ensemble density of particles.
\par
A simple way to prove (\ref{aa}), is to make
the canonical transformation $\psi =\sqrt{\rho}\exp(iS/\hbar)$ in the action
for the Schr\"odinger equation. The equations of motion of $\rho$ and $S$
would be:
\begin{equation}
\frac{\partial \rho}{\partial t}+\vec{\nabla}\cdot \left ( \rho \frac{\vec{\nabla}S}{m}\right )=0
\label{bb}
\end{equation}
\begin{equation}
\frac{\partial S}{\partial t}+\frac{|\vec{\nabla}S|^2}{2m}+V+Q=0
\label{dd}
\end{equation}
in which $V$ is the classical potential and $S$ is the Hamilton--Jacobi function. Equation (\ref{bb}) is the continuity equation
provided (\ref{cc}) is satisfied. Under this condition the quantum Hamilton-Jacobi equation (\ref{dd})
is identical to (\ref{aa}).
\subsection{Relativistic de-Broglie--Bohm theory}
Extension of de-Broglie--Bohm causal theory of quantum phenomena to the relativistic case is a problematic matter. Essentially all the problems of the Copenhagen relativistic quantum mechanics can in principle be present in the de-Broglie--Bohm theory. There are at least three problems with Copenhagen relativistic quantum mechanics. They are:
\begin{itemize}
\item The problem of negative energy and space--like current densities for integer spins.
\item The problem of defining probability distribution for many--particle systems.
\item The conflict between measurement principle (which states that measurement is instantaneous) and the Poincar\`e transformations.
\end{itemize}
We shall discus about the second problem at the end of this paper. The third one is essentially not a problem in de-Broglie--Bohm theory, provided the second problem is solved. This is because of the fact that measurement is not an instantaneous phenomena in this theory, it is a many--particle situation.

In a recent paper\cite{rel} we have shown that the first problem is not present in de-Broglie--Bohm theory, provided one sets the natural constraint that the theory should lead the correct non--relativistic limit. Let us make this point clear. Usually one gets a de-Broglie--Bohm version of a Copenhagen theory by writing the wave function in its polar form $\psi=|\psi|\exp(iS/\hbar)$ and decomposing the real and imaginary parts of the wave equation. Doing this with the Klein--Gordon equation leads to a quantum Hamilton--Jacobi equation:
\begin{equation}
\partial_\mu S \partial^\mu S=m^2c^2(1+Q)
\label{ee}
\end{equation}
with the quantum potential defined as:
\begin{equation}
Q=\frac{\hbar^2}{m^2c^2}\frac{\Box|\psi|}{|\psi|}
\end{equation}
and the continuity equation:
\begin{equation}
\partial_\mu(\rho\partial^\mu S)=0
\end{equation}
The above Hamilton--Jacobi equation (\ref{ee}) shows that in the relativistic case the quantum potential is essentially the mass square. So one can define the quantum mass of a particle as:
\begin{equation}
{\cal M}^2=m^2(1+Q)
\end{equation}
Since the quantum potential can be a negative number, in general the tachyonic solutions would emerge. This is essentially related to the first problem noted above. Although it can be shown that a non--tachyonic initial condition leads to a global (in time) non--tachyonic solution\cite{deb}, but the existance of tachyonic solutions is a fatal problem.

It can be shown that the problem is that equation (\ref{ee}) is not the correct relativistic equation of motion\cite{rel}. A correct relativistic quantum equation of motion \textit{should not only be poincar\`e invariant but also has the correct non--relativistic limit}. In \cite{rel} we have shown that using these requirements one gets the correct equation of motion as:
\begin{equation}
\partial_\mu S\partial^\mu S ={\cal M}^2c^2
\end{equation}
with
\begin{equation}
{\cal M}^2=m^2\exp\left ( Q\right )
\end{equation}
this clearly is free from the above mentioned problem.
\subsection{de-Broglie--Bohm theory in curved space--time}
The extension to the case of a particle moving in a curved background is not very difficult\cite{bohm}. This can be done through the same way as writing any special relativistic relation in a general relativistic form. One should only change the ordinary differentiating $\partial_\mu$ with the covariant derivative $\nabla_\mu$ and change the Lorentz metric $\eta_{\mu\nu}$ to the curved metric $g_{\mu\nu}$.

Therefore the equations of motion for a particle (of spin zero) in a curved background are:
\begin{equation}
\nabla_\mu\left (\rho\nabla^\mu S\right )=0
\end{equation}
\begin{equation}
g^{\mu\nu}\nabla_\mu S \nabla_\nu S={\cal M}^2c^2
\label{a}
\end{equation}
where
\begin{equation}
{\cal M}^2=m^2\exp\left (Q\right )
\label{d}
\end{equation}
\begin{equation}
Q=\frac{\hbar^2}{m^2c^2}\frac{\Box_g|\psi|}{|\psi|}
\end{equation}

de-Broglie made the following interesting and
fruitfull observation\cite{deb}: The quantum Hamilton-Jacobi equation (\ref{a}) can be written as:
\begin{equation}
\frac{m^2}{{\cal M}^2}g^{\mu \nu}\nabla_{\mu}S\nabla_{\nu}S=m^2c^2
\label{b}
\end{equation}
From this relation it can be concluded
that the quantum effects are identical with the change of the space-time metric from
$g_{\mu \nu}$ to:
\begin{equation}
g_{\mu \nu}\longrightarrow\widetilde{g}_{\mu\nu}=\frac{{\cal M}^2}{m^2}g_{\mu \nu}
\label{c}
\end{equation}
which is a conformal transformation.
\par
Therefore equation (\ref{b}) can be written as:
\begin{equation}
\widetilde{g}^{\mu\nu}\widetilde{\nabla}_{\mu}S\widetilde{\nabla}_{\nu}S=m^2c^2
\end{equation}
where $\widetilde{\nabla}_{\mu}$ represents the covariant
differentiation with respect to the metric $\widetilde{g}_{\mu \nu}$.
In this new curved space-time the other equation of motion, i.e. the continuity
relation should be written as:
\begin{equation}
\widetilde{g}^{\mu\nu}\widetilde{\nabla} _{\mu}\left ( \rho \widetilde{\nabla}_{\nu}S \right )=0
\end{equation}

The important conclusion we draw from this argumentation is that
\textit{ the presence of the quantum
potential is equivalent to a curved space-time with its metric being given by (\ref{c})}. So in fact, we have the geometrization of the quantum
aspects of matter.
In this way,
it seems that there is a dual aspect to the role of geometry in physics.
The space-time geometry sometimes looks like what we call gravity and sometimes
looks like what we understand as quantal behaviours. Since
the equations governing the space-time geometry are highly non-linear, the
curvature due to the quantum potential
may have a large influence on the classical contribution to the
curvature of the space-time. This would be investigated in the following sections.

The particle trajectory can be derived from the guidance relation and  by differentiating (\ref{a}) leading to  Newton's equation of motion:
\begin{equation}
\label{qg}
{\cal M}\frac{d^2x^\mu}{d\tau^2}+{\cal M}\Gamma^\mu_{\nu\kappa}u^\nu u^\kappa
=(c^2g^{\mu\nu}-u^\mu u^\nu)\nabla_\nu{\cal M}\,.
\label{m}
\end{equation}
using the above  conformal transformation, Eq.(\ref{m}) reduces to the standard
geodesic equation via the above conformal transformation.
\section{A tensor model of the idea}
\subsection[The case of an ensemble of particles]{The case of an ensemble of particles\cite{geo}\label{s1}}
A general relativistic system consisting of gravity and classical matter (relativistic particles without quantum effects) is determined by the action:
\begin{equation}
{\cal A}_{no-quantum}=\frac{1}{2\kappa}\int d^4x \sqrt{-g}{\cal R}+\int
d^4x \sqrt{-g}\frac{\hbar^2}{m}
\left ( \frac{\rho}{\hbar^2}{\cal D}_{\mu}S{\cal D}^{\mu}S-\frac{m^2}{\hbar^2}\rho \right )
\label{ff}
\end{equation}
where $\kappa=8\pi G$ and hereafter we chose the units in which $c=1$.

On the other hand as it was seen in the previous section, the de-Broglie remark leads
to the conclusion that introducing  the quantum potential, is equivalent to the introduction of a conformal factor
$\Omega^2={\cal M}^2/m^2$
in the metric. So in order to introduce the rather than quantum effects of matter into
the action (\ref{ff}), we make the aforementioned conformal transformation, instead of
adding the quantum potential term.

Accordingly, we write our action with quantum effects as:
\[ {\cal A}[\overline{g}_{\mu \nu},\Omega,S,\rho,\lambda]=\frac{1}{2\kappa}\int d^4x\sqrt{-\overline{g}}\left (\overline{{\cal R}}\Omega^2-6\overline{\nabla}_{\mu}\Omega\overline{\nabla}^{\mu}\Omega\right )\]
\begin{equation}
+\int d^4x \sqrt{-\overline{g}}\left ( \frac{\rho}{m}\Omega^2\overline{\nabla}_{\mu}S\overline{\nabla}^{\mu}S -m\rho\Omega^4\right)
+\int d^4x \sqrt{-\overline{g}}\lambda \left [ \Omega^2-\left ( 1+\frac{\hbar^2}{m^2}\frac{\overline{\Box}\sqrt{\rho}}{\sqrt{\rho}}\right ) \right ]
\end{equation}
where a bar over any quantity means that it corresponds to no--quantum regime.
Here we used only the first two terms of expansion of equation (\ref{d}) to kip things simple. No physical change emerges considering all terms.
In the above action, $\lambda$ is a Lagrange multiplier which is introduced to identify
the conformal factor with its Bohmian value.

Here two problems must be noted. First, in the above action, we use
$\overline{g}_{\mu \nu}$ to raise or
lower indices and to evaluate the covariant derivatives. Second, the
physical metric (i.e. the metric containing the quantum effects of matter) is
$g_{\mu \nu}$ given by $\Omega^2\overline{g}_{\mu \nu}$.

By the variation of the above action  with respect to $\overline{g}_{\mu \nu}$, $\Omega$, $\rho$, $S$ and $\lambda$
we arrive at the following relations as our quantum  equations of motion:
\begin{enumerate}
\item The equation of motion for $\Omega$:
\begin{equation}
\overline{{\cal R}}\Omega+6\overline{\Box}\Omega+\frac{2\kappa}{m}\rho \Omega \left ( \overline{\nabla}_{\mu}S\overline{\nabla}^{\mu}S-2m^2\Omega^2\right )+2\kappa \lambda\Omega=0
\label{gg}
\end{equation}
\item The continuity equation for particles:
\begin{equation}
\overline{\nabla}_{\mu}\left (\rho \Omega^2 \overline{\nabla}^{\mu}S \right )=0
\end{equation}
\item The equation of motion for particles:
\begin{equation}
\left ( \overline{\nabla}_{\mu}S \overline{\nabla}^{\mu}S -m^2\Omega^2\right )\Omega^2\sqrt{\rho}+\frac{\hbar^2}{2m}\left [ \overline{\Box}\left (\frac{\lambda}{\sqrt{\rho}}\right ) -\lambda\frac{\overline{\Box}\sqrt{\rho}}{\rho}\right ]=0
\end{equation}
\item The modified Einstein equations for $\overline{g}_{\mu \nu}$:
\[ \Omega^2\left [ \overline{{\cal R}}_{\mu \nu}-\frac{1}{2}\overline{g}_{\mu \nu}\overline{{\cal R}}\right ]
-\left [ \overline{g}_{\mu \nu}\overline{\Box} -\overline{\nabla}_{\mu}\overline{\nabla}_{\nu}\right ]
\Omega^2 -6 \overline{\nabla}_{\mu}\Omega \overline{\nabla}_{\nu}\Omega+3\overline{g}_{\mu \nu}\overline{\nabla}_{\alpha}\Omega \overline{\nabla}^{\alpha}\Omega\]
\[ +\frac{2\kappa}{m}\rho\Omega^2 \overline{\nabla}_{\mu}S \overline{\nabla}_{\nu}S-\frac{\kappa}{m}\rho \Omega^2\overline{g}_{\mu \nu} \overline{\nabla}_{\alpha}S \overline{\nabla}^{\alpha}S +\kappa m \rho \Omega^4 \overline{g}_{\mu \nu}\]
\begin{equation}
+\frac{\kappa\hbar^2}{m^2}\left [ \overline{\nabla}_{\mu}\sqrt{\rho}\overline{\nabla}_{\nu}\left ( \frac{\lambda}{\sqrt{\rho}}\right )
 +\overline{\nabla}_{\nu}\sqrt{\rho}\overline{\nabla}_{\mu}\left ( \frac{\lambda}{\sqrt{\rho}}\right )\right]
 -\frac{\kappa\hbar^2}{m^2}\overline{g}_{\mu \nu}\overline{\nabla}_{\alpha}\left [\lambda \frac{\overline{\nabla}^{\alpha}\sqrt{\rho}}{\sqrt{\rho}}\right ]=0
\label{hh}
\end{equation}
\item The constraint equation:
\begin{equation}
\Omega^2=1+\frac{\hbar^2}{m^2}\frac{\overline{\Box}\sqrt{\rho}}{\sqrt{\rho}}
\label{v}
\end{equation}
\end{enumerate}
\par
As it is seen, the back--reaction effects of the quantum factor on the background
metric are contained in those highly coupled equations. It may be
noted that by combining (\ref{gg}) and (\ref{hh}) it is possible to arrive at
a more simple relation instead of (\ref{gg}). If we take the trace of (\ref{hh}) and
use (\ref{gg}),
we have after some mathematical manipulations:
\begin{equation}
\lambda=\frac{\hbar^2}{m^2}\overline{\nabla}_{\mu}\left [ \lambda \frac{\overline{\nabla}^{\mu}\sqrt{\rho}}{\sqrt{\rho}}\right ]
\label{ii}
\end{equation}

Before proceeding, some important point about this relation must be noted. If one tries
to solve it via perturbation method, in terms of the small parameter:
\begin{equation}
\alpha=\frac{\hbar^2}{m^2}
\end{equation}
by writting:
\begin{equation}
\lambda=\lambda^{(0)}+\alpha\lambda^{(1)}+\alpha^2\lambda^{(2)}+\cdots
\end{equation}
and
\begin{equation}
\sqrt{\rho}=\sqrt{\rho}^{(0)}+\alpha\sqrt{\rho}^{(1)}+\alpha^2\sqrt{\rho}^{(2)}+\cdots
\end{equation}
one gets:
\begin{equation}
\lambda^{(0)}=\lambda^{(1)}=\lambda^{(2)}=\cdots=0
\end{equation}
So the perturbative solution of (\ref{ii}) is $\lambda=0$ which is its
trivial solution.

Therefore, our equations are:
\begin{equation}
\overline{\nabla}_{\mu}\left (\rho \Omega^2 \overline{\nabla}^{\mu}S \right )=0
\label{w}
\end{equation}
\begin{equation}
\overline{\nabla}_{\mu}S \overline{\nabla}^{\mu}S =m^2\Omega^2
\label{u}
\end{equation}
\begin{equation}
{\cal G}_{\mu \nu}=-\kappa {\cal T}^{(m)}_{\mu\nu}-\kappa{\cal T}^{(\Omega)}_{\mu\nu}
\label{x}
\end{equation}
where ${\cal T}^{(m)}_{\mu\nu}$ is the matter energy--momentum tensor and
\begin{equation}
\kappa{\cal T}^{(\Omega)}_{\mu\nu}=\frac{\left [ g_{\mu \nu}\Box
-\nabla_{\mu}\nabla_{\nu}\right ]
\Omega^2}{\Omega^2} +6 \frac{\nabla_{\mu}\Omega \nabla_{\nu}\Omega}{\omega^2}
-3g_{\mu \nu}\frac{\nabla_{\alpha}\Omega \nabla^{\alpha}\Omega}{\Omega^2}
\label{y}
\end{equation}
and
\begin{equation}
\Omega^2=1+\alpha\frac{\overline{\Box}\sqrt{\rho}}{\sqrt{\rho}}
\end{equation}
Note that relation (\ref{u}) is, in fact, the Bohmian equation of motion, and if
we write
it in terms of the physical metric $g_{\mu \nu}$, it reads as $\nabla_{\mu}S\nabla^{\mu}S=m^2c^2$.
This is what we expect from de-Broglie's conjecture.
\subsection[The case of a single particle]{The case of a single particle\cite{qgg}\label{s2}}
In the previous subsection we have assumed that there is a real ensemble of the quantum particle. Now the question is what happens for the case of a single particle? To investigate this, we first examine how we can translate the quantum
potential in a complete geometrical manner, i.e.~we write it in a form
that there is no explicit reference to matter parameters. Only after
using the field equations can one deduce the original form of the
quantum potential. This has the advantage of allowing our theory to
work both for a single particle and an ensemble. Next, we write a
special field equation as a toy theory and extract some of its
consequences.
\subsubsection{Geometry of the quantum conformal factor}
Let us first ignore gravity and examine the geometrical properties of the conformal factor
given~by
\begin{equation}
g_{\mu\nu}=e^{4\Sigma}\eta_{\mu\nu};\qquad e^{4\Sigma}=\frac{{\cal M}^2}{m^2} =\exp\left (\alpha\frac{\square_\eta\sqrt{\rho}}{\sqrt{\rho}}\right )
=\exp\left (\alpha\frac{\square_\eta\sqrt{|{\cal T}|}}{\sqrt{|{\cal T}|}}\right )\,,
\label{jj}
\end{equation}
where ${\cal T}$ is the trace of the energy--momentum tensor\footnote{The absolute value sign is introduced to
make the square root always meaningful.} and is substituted for $\rho$ (as it is true for dust).

Evaluating the Einstein's tensor for the above metric, we have:
\begin{equation}
{\cal G}_{\mu\nu}=4g_{\mu\nu}e^\Sigma\square_\eta e^{-\Sigma}+2e^{-2\Sigma}\partial_\mu\partial_\nu e^{2\Sigma}
\end{equation}
So as an ansatz, we suppose that in the presense of gravitational effects,
the field equation have some form like:
\begin{equation}
{\cal R}_{\mu\nu}-\frac{1}{2}{\cal R}g_{\mu\nu} = \kappa {\cal T}_{\mu\nu} + 4 g_{\mu\nu} e^\Sigma \square e^{-\Sigma} + 2 e^{-2\Sigma} \nabla_\mu \nabla_\nu e^{2\Sigma}\,.
\label{j}
\end{equation}
This equation is written in such a way that in the limit ${\cal
T}_{\mu\nu}\rightarrow 0$ the solution (\ref{jj}) achieved.

Making the trace of the above equation one gets
\begin{equation}
-{\cal R}=\kappa {\cal T}-12\square\Sigma+24(\nabla\Sigma)^2
\end{equation}
which has the iterative solution:
\begin{equation}
\kappa{\cal T}=-{\cal R}+12\alpha\square\left( \frac{\square
\sqrt{{\cal R}}}{\sqrt{{\cal R}}}\right) +\cdots
\end{equation}
leading to
\begin{equation}
\Sigma=\alpha \frac{\square \sqrt{|{\cal T}|}}{\sqrt{|{\cal T}|}}\simeq \alpha
\frac{\square \sqrt{|{\cal R}|}}{\sqrt{|{\cal R}|}}
\end{equation}
up to first order in $\alpha$. Now we are ready to make a toy model.
\subsubsection{Field equations of a toy quantum gravity\label{s14}}
From the above equation we learn that ${\cal T}$ can be
replaced with ${\cal R}$ in the expression for the quantum potential
or for the conformal factor of the space--time metric. This
replacement is in fact an important improvement, because the explicit
reference to ensemble density is removed. This allows the theory to work
for both a single particle and an ensemble.

So with a glance at Eq.~(\ref{j}) for our toy quantum--gravity
theory, we assume the following field equations:
\begin{equation}
{\cal G}_{\mu\nu}-\kappa{\cal T}_{\mu\nu}-{\cal Z}_{\mu\nu\alpha\beta}
\exp\left[\frac{\alpha}{2}\Phi\right]
\nabla^\alpha\nabla^\beta
\exp\left[-\frac{\alpha}{2}\Phi\right]
=0\,,
\label{l}
\end{equation}
where
\begin{equation}
{\cal Z}_{\mu\nu\alpha\beta}=2\left[ g_{\mu\nu}g_{\alpha\beta}-g_{\mu\alpha}
g_{\nu\beta}\right]
\end{equation}
\begin{equation}
\Phi=\frac{\square\sqrt{|{\cal R}|}}{\sqrt{|{\cal R}|}}\,.
\end{equation}
Note that the number $2$ and the minus sign of the second term of the
last equation are chosen so that the energy equation derived later
be correct.  It would be very useful to take the trace of
Eq.~(\ref{l}):
\begin{equation}
{\cal R}+\kappa{\cal T}+6\exp\left[{\alpha\Phi\over
2}\right]\square\exp\left[-{\alpha\Phi\over 2}\right]=0\,.
\label{k}
\end{equation}
In fact this equation represents the connection of Ricci scalar
curvature of  space--time and the trace of matter energy--momentum
tensor.  In the cases when a perturbative solution is
admitted, i.e.~when we \textit{can} expand anything in terms of powers
of $\alpha$, one can find the relation between ${\cal R}$ and ${\cal
T}$ perturbatively. In the zeroth approximation one has the classical
relation:
\begin{equation}
{\cal R}^{(0)}=-\kappa{\cal T}\,.
\end{equation}
As a better approaximation up to first order in $\alpha$, one gets
\begin{equation}
{\cal R}^{(1)}=-\kappa{\cal T}
-6\exp\left[{\alpha\Phi^{(0)}\over 2}\right]\square\exp\left[-{\alpha\Phi^{(0)}\over 2}\right]\,,
\end{equation}
where
\begin{equation}
\Phi^{(0)}=\frac{\square\sqrt{|{\cal T}|}}{\sqrt{|{\cal T}|}}\,.
\end{equation}
A better result can be obtained in the second order as
\begin{eqnarray}
{\cal R}^{(2)}&=&-\kappa{\cal T}
-6\exp\left[{\alpha\Phi^{(0)}\over 2}\right]\square\exp\left[-{\alpha\Phi^{(0)}\over 2}\right]\nonumber\\[6pt]
&&{}-6\exp\left[{\alpha\Phi^{(1)}\over 2}\right]\square\exp\left[-{\alpha\Phi^{(1)}\over 2}\right]
\end{eqnarray}
with
\begin{equation}
\Phi^{(1)}=\frac{\square\sqrt{|-\kappa{\cal T}-6
\exp[\alpha\Phi^{(0)}/2]\square\exp[-\alpha\Phi^{(0)}/2]
|}}{\sqrt{|-\kappa{\cal T}
\exp[\alpha\Phi^{(0)}/2]\square\exp[-\alpha\Phi^{(0)}/2]
|}}\,.
\end{equation}

The energy relation can be obtained via taking the four divergence of
the field equations. Using the fact that the divergence of
Einstein's tensor is zero, one gets
\begin{equation}
\kappa\nabla^\nu{\cal T}_{\mu\nu}=\alpha{\cal R}_{\mu\nu}\nabla^\nu\Phi-
\frac{\alpha^2}{4} \nabla_\mu(\nabla\Phi)^2+\frac{\alpha^2}{2}\nabla_\mu\Phi\square\Phi\,.
\end{equation}
For a dust with
\begin{equation}
{\cal T}_{\mu\nu}=\rho u_\mu u_\nu
\label{aaa}
\end{equation}
where $u_\mu$ is the velocity field. Assuming the conservation law for mass
\begin{equation}
\nabla^\nu\left( \rho{\cal M}u_\nu\right)=0
\end{equation}
up to first order in $\alpha$ one arrives at:
\begin{equation}
\frac{\nabla_\mu{\cal M}}{{\cal M}}=-\frac{\alpha}{2}\nabla_\mu\Phi
\end{equation}
or
\begin{equation}
{\cal M}^2=m^2\exp (-\alpha\Phi)\,,
\end{equation}
where $m$ is some integration constant. This is the correct relation of mass field and the quantum potential.
\section{A scalar--tensor model of the idea}
In the last section, the form of the quantum potential and its relation to the conformal degree of freedom of the space--time metric are assumed. The next step is to remove these assumptions and \textit{derive} them using the equations of motion. In doing this we first include the conformal factor as a scalar field and then introduce another scalar field (quantum potential in fact). On the equations of motion the correct relation between quantum potential and conformal factor and also the form of the quantum potential would emerge.
\subsection[Making the conformal factor dynamical]{Making the conformal factor dynamical\cite{st}\label{s3}}
We start from the most general scalar--tensor action:
\begin{equation}
{\cal A}=\int d^4x \left \{ \phi {\cal R}-\frac{\omega}{\phi}\nabla^\mu\phi
\nabla_\mu\phi+2\Lambda\phi+{\cal L}_m\right \}
\end{equation}
in which $\omega$ is a constant independent of the scalar field, and $\Lambda$
is the cosmological constant. Also, it is
assumed that the matter lagrangian is coupled to the scalar field.
The equations of motion are:
\begin{equation}
{\cal R}+\frac{2\omega}{\phi}\Box\phi-\frac{\omega}{\phi^2}\nabla^\mu\phi
\nabla_\mu\phi+2\Lambda+\frac{\partial {\cal L}_m}{\partial\phi}=0
\label{kk}
\end{equation}
\begin{equation}
{\cal G}^{\mu\nu}-\Lambda g^{\mu\nu}=-\frac{1}{\phi}{\cal T}^{\mu\nu}
-\frac{1}{\phi}[\nabla^\mu\nabla^\nu-g^{\mu\nu}\Box ]\phi+\frac{\omega}{\phi^2}
\nabla^\mu\phi\nabla^\nu\phi-\frac{1}{2}\frac{\omega}{\phi^2}g^{\mu\nu}
\nabla^\alpha\phi\nabla_\alpha\phi
\end{equation}

The scalar curvature can be evaluated from the contracted form of the latter
equation, and it can be substituted in the relation (\ref{kk}). Then we have:
\begin{equation}
\frac{2\omega-3}{\phi}\Box\phi=-\frac{{\cal T}}{\phi}+2\Lambda-
\frac{\partial{\cal L}_m}{\partial\phi}
\label{ll}
\end{equation}
The matter lagrangian for an ensemble of relativistic particles of mass $m$
is (without any quantum contribution):
\begin{equation}
{\cal L}_{m(no-quantum)}=\frac{\rho}{m}\nabla_\mu S\nabla^\mu S-\rho m
\end{equation}
This lagrangian can be generalized if one assumes that
there is some interaction between the scalar field and the matter field.
Here, for simplicity, it is assumed that this interaction is in the form of
powers of $\phi$.
In order to bring in the quantum effects, one needs to add terms containig
the quantum potential.
Physical intuition leads us to the fact that it is
necessary to assume some interaction between cosmological constant and
matter quantum potential. This suggestion will be confirmed after
obtaining all of the equations of motion.
These arguments lead us to consider the matter lagrangian as:
\begin{equation}
{\cal L}_m=\frac{\rho}{m}\phi^a\nabla^\mu S\nabla_\mu S-m\rho\phi^b-\Lambda(1+Q)^c
\end{equation}
in which the $a$, $b$, and $c$ constants have to be fixed later.  Again we used the first two terms of equation (\ref{d}) for simplicity. Therefore the
energy--momentum tensor is:
\[ {\cal T}^{\mu\nu}=-\frac{1}{\sqrt{-g}}\frac{\delta}{\delta g_{\mu\nu}}
\int d^4x \sqrt{-g} {\cal L}_m=-\frac{1}{2}g^{\mu\nu}{\cal L}_m+\frac{\rho}{m}
\phi^a\nabla^\mu S\nabla^\nu S -\frac{1}{2}\Lambda cQ(1+Q)^{c-1}g^{\mu\nu}\]
\begin{equation}
-\frac{1}{2}\alpha\Lambda c\nabla_\alpha\sqrt{\rho}\nabla_\beta\left (
\frac{(1+Q)^{c-1}}{\sqrt{\rho}}\right )\left [ g^{\mu\nu}g^{\alpha\beta}
-g^{\alpha\mu}g^{\beta\nu}-g^{\beta\mu}g^{\alpha\nu}\right ]
\end{equation}
Using the matter lagrangian and contracting the above tensor, one can
calculate the first and third terms in the relation (\ref{ll}).
The other equations, the continuity equation and the quantum Hamilton--Jacobi
equation, are expressed respectively as:
\begin{equation}
\nabla^\mu\left ( \rho\phi^a\nabla_\mu S\right )=0
\end{equation}
\begin{equation}
\nabla^\mu S \nabla_\mu S=m^2\phi^{b-a}-\frac{1}{2}\Lambda mc\frac{Q}{\rho\phi^a}
(1+Q)^{c-1}+\frac{1}{2}\Lambda mc\alpha\frac{1}{\sqrt{\rho}\phi^a}
\Box\left ( \frac{(1+Q)^{c-1}}{\sqrt{\rho}}\right )
\end{equation}
To simplify the calculations, with due attention to the
equation (\ref{ll}), one can choose $\omega$ to be $\frac{3}{2}$. Then a perturbative
expansion for the scalar field and matter distribution density can be used as:
\begin{equation}
\phi=\phi_0+\alpha\phi_1+\cdots
\end{equation}
\begin{equation}
\sqrt{\rho}=\sqrt{\rho_0}+\alpha\sqrt{\rho_1}\cdots
\end{equation}
In the zeroth order approaximation, the scalar field equation gives:
\begin{equation}
b=a+1;\ \ \ \ \ \ \ \ \phi_0=1
\end{equation}
In the first order approaximation one gets:
\begin{equation}
\alpha\phi_1=\frac{c}{2}(1-a)Q+\frac{a}{2}c\tilde{Q}
\end{equation}
in which:
\begin{equation}
\tilde{Q}=\alpha\frac{\nabla_\mu\sqrt{\rho}\nabla^\mu\sqrt{\rho}}{\rho}
\end{equation}

Since the scalar field is the conformal factor
of the space--time metric, and because of some arguments\cite{geo,conf} show
that this field is a function of matter quantum potential, one might
choose the constant $a$ equall to zero. Then, the scalar field
is independent of $\tilde{Q}$ and we have:
\begin{equation}
\alpha\phi_1=\frac{c}{2}Q
\end{equation}
Also the Bohmian equations of motion give:
\begin{equation}
\nabla_\mu S\nabla^\mu S=m^2(1+cQ/2)-\Lambda mc\frac{Q-\tilde{Q}}{\rho_0}
\end{equation}
It is necessary to choose $c=2$ in order that the first term on the right
hand side be the same as the quantum mass ${\cal M}$.
These choises for parameters $a$, $b$ and $c$ lead to
the non--perturbative quantum gravity equations
as follows:
\begin{equation}
\phi=1+Q-\frac{\alpha}{2}\Box Q
\label{nn}
\end{equation}
\begin{equation}
\nabla^\mu S\nabla_\mu S=m^2\phi-\frac{2\Lambda m}{\rho}(1+Q)(Q-\tilde{Q})
+\frac{\alpha\Lambda m}{\rho}\left ( \Box Q -2\nabla_\mu Q\frac{\nabla^\mu
\sqrt{\rho}}{\sqrt{\rho}}\right )
\label{oo}
\end{equation}
\begin{equation}
\nabla^\mu(\rho\nabla_\mu S)=0
\end{equation}
\begin{equation}
{\cal G}^{\mu\nu}-\Lambda g^{\mu\nu}=-\frac{1}{\phi}{\cal T}^{\mu\nu}
-\frac{1}{\phi}[\nabla^\mu\nabla^\nu-g^{\mu\nu}\Box ]\phi+\frac{\omega}{\phi^2}
\nabla^\mu\phi\nabla^\nu\phi-\frac{1}{2}\frac{\omega}{\phi^2}g^{\mu\nu}
\nabla^\alpha\phi\nabla_\alpha\phi
\label{mm}
\end{equation}

We conclude this section by pointing out some important hints:
\begin{itemize}
\item It is very interesting that in the framework of the scalar--tensor
theories, one is able to derive all of the quantum gravity equations of motion
without using the method of lagrange.
\item In the suggested quantum gravity theory, the causal structure of the
space--time ($g_{\mu\nu}$) is determined via equation (\ref{mm}). This
shows that except for back--reaction terms of the quantum effects on $g_{\mu\nu}$,
the causal structure of the space--time is determined by the gravitational
effects of matter. Quantum effects, determine directly the scale factor
of the space--time, from the relation (\ref{nn}).
\item It must be noted that the mass field given by the right hand side
of the relation (\ref{oo}), consists of two parts. The first part which
is proportionnal to $\alpha$, is a purely quantum effect, and the second part
which is proportional to $\alpha\Lambda$, is a mixture of the quantum effects
and the large scale structure introduced via the cosmological constant.
\item In the present theory, the scalar field produces quantum force that appears on right hand side and violates the equivalence principle. Similarly, in
Kalutza--Klein theory, the scalar field (dilaton) produces some fifth force leading to the violation of the equivalence principle\cite{cho}.
\end{itemize}
\subsection[Making the quantum potential dynamical]{Making the quantum potential dynamical\cite{ss}\label{s4}}
Using the findings of the previous subsection, one can write an appropriate action such that
the conformal factor and quantum potential are both dynamical
fields. In this way, the relation between the conformal factor and
quantum potential, and also the dependence of quantum potential to the
ensemble density are resulted at the first order of approximation.  In this way one deals with a scalar--tensor theory with two scalar fields. Thus we start from the most general action:
\begin{equation}	
{\cal A}=\int d^4x\ \sqrt{-g}
\left[ \phi{\cal R}-\omega
\frac{\nabla_\mu\phi\nabla^\mu\phi}{\phi}-
\frac{\nabla_\mu Q\nabla^\mu Q}{\phi}+2\Lambda\phi+{\cal L}_m\right]\,,
\end{equation}
The cosmological constant generally has an
interaction term with the scalar field.  We prefer to use the matter
Lagrangian:
\begin{equation}	
{\cal L}_m=\frac{\rho}{m}\phi^a\nabla_\mu S\nabla^\mu S-m
\rho\phi^b-\Lambda (1+Q)^c+\alpha\rho (e^{\beta Q}-1)\,.
\end{equation}
The first three terms of this Lagrangian are the same as those of the
previous subsection. The last term is chosen in such a way, that
satisfies two facts. It is necessary to have an interaction between
the quantum potential field and the ensemble density, to have a
relation between them via the equations of motion.  Furthermore, this
interaction is written such that in the classical limit, it vanishes.

Variation of the above action functional leads to the following
equations \hbox{of~motion}:

\begin{itemize}
\item the scalar field's equation of motion
\begin{eqnarray}	
&&{\cal R}+\frac{2\omega}{\phi}\square\phi-\frac{\omega}{\phi^2}
\nabla^\mu\phi\nabla_\mu\phi +2\Lambda \nonumber \\[12pt]
&&\hskip20pt{}+\frac{1}{\phi^2}\nabla^\mu Q\nabla_\mu Q+
\frac{a}{m}\rho\phi^{a-1}\nabla^\mu S
\nabla_\mu S-mb\rho\phi^{b-1}=0 \label{pp}
\end{eqnarray}

\item the quantum potential's equation of motion
\begin{equation}	
\frac{\square Q}{\phi}-\frac{\nabla_\mu Q\nabla^\mu\phi}{\phi^2}-
\Lambda c(1+Q)^{c-1}+\alpha\beta\rho e^{\beta Q}=0 \label{ss}
\end{equation}

\item the generalized Einstein's equation
\begin{eqnarray}	
{\cal G}^{\mu\nu}-\Lambda g^{\mu\nu}&=&-\frac{1}{\phi}
{\cal T}^{\mu\nu}-\frac{1}{\phi}
\left[ \nabla^\mu\nabla^\nu-g^{\mu\nu}\square\right] \phi+
\frac{\omega}{\phi^2} \nabla^\mu\phi\nabla^\nu\phi \nonumber \\[12pt]
&&{}-\frac{\omega}{2\phi^2}g^{\mu\nu}\nabla^\alpha\phi
\nabla_\alpha\phi +\frac{1}{\phi^2}\nabla^\mu Q\nabla^\nu Q-
\frac{1}{2\phi^2} g^{\mu\nu}\nabla^\alpha Q\nabla_\alpha Q\qquad
\label{qq}
\end{eqnarray}

\item the continuity equation
\begin{equation}	
\nabla_\mu (\rho\phi^a\nabla^\mu S)=0
\end{equation}

\item the quantum Hamilton--Jacobi equation
\begin{equation}	
\nabla^\mu S\nabla_\mu S=m^2\phi^{b-a}-\alpha m\phi^{-a}
(e^{\beta Q}-1)\,.\label{rr}
\end{equation}
\end{itemize}

\noindent
In Eq.~(\ref{pp}), the scalar curvature and the term $\nabla^\mu
S\nabla_\mu S$ can be eliminated using Eqs.~(\ref{qq}) and
(\ref{rr}). In addition, on using the matter Lagrangian and the
definition of the energy--momentum tensor, one has
\begin{equation}	
(2\omega-3) \square\phi=(a+1)\rho\alpha (e^{\beta Q}-1)-2
\Lambda(1+Q)^c+2\Lambda\phi-\frac{2}{\phi}
\nabla_\mu Q\nabla^\mu Q\,, \label{tt}
\end{equation}
where the constant $b$ is chosen as $a+1$ as in the previous subsection. We solve the Eqs.~(\ref{ss}) and (\ref{tt}),
using perturbative expansion with $\alpha$ as the \hbox{expansion~parameter}:
\begin{eqnarray}	
Q&=&Q_0+\alpha Q_1+\cdots \,, \\[8pt]
\phi &=&1+\alpha Q_1+\cdots \,, \\[8pt]
\sqrt{\rho}&=&\sqrt{\rho}_0+\alpha\sqrt{\rho}_1+\cdots\,,
\end{eqnarray}
where the conformal factor is chosen to be unity at the zeroth order
of perturbation, because in the limit $\alpha\rightarrow 0$
Eq.~(\ref{rr}) would lead to the classical Hamilton--Jacobi equation.
Since by Eq.~(\ref{rr}), the quantum mass is given by $m^2\phi+{\rm
other\ terms}$, the first order term of $\phi$ is chosen to be $Q_1$
as it must be so according to the relation of quantum mass
(\ref{d}). Also we shall show that $Q_1$ would be equal to
$\square\sqrt{\rho}/\sqrt{\rho}$ plus some corrections, which is
desired as we called $Q$ the quantum potential field.

At the zeroth order one gets
\begin{eqnarray}	
\square Q_0-\Lambda c -\Lambda c(c-1)Q_0&=&0\,, \label{uu} \\[8pt]
\nabla_\mu Q_0\nabla^\mu Q_0&=&-\Lambda cQ_0 \label{i}
\end{eqnarray}
and at the first order:
\begin{eqnarray}	
\nabla_\mu Q_0\nabla^\mu Q_1&=&\square Q_1-Q_1\square Q_0-
\Lambda c(c-1)Q_1+\beta\rho_0 e^{\beta Q_0} \label{vv} \\[8pt]
(2\omega-3)\square Q_1&=&(a+1)\rho_0 (e^{\beta Q_0}-1)
-2\Lambda (c-1) Q_1 \nonumber \\[8pt]
&&{}-4\nabla_\mu Q_0\nabla^\mu Q_1+2Q_1\nabla_\mu Q_0
\nabla^\mu Q_0\,. \label{ww}
\end{eqnarray}
On using Eqs.~(\ref{uu}), (\ref{i}) and (\ref{vv}), in
Eq.~(\ref{ww}), one get:
\begin{equation}	
\square Q_1+A (\rho_0)Q_1+B(\rho_0)=0\,,\label{e}
\end{equation}
where
\begin{eqnarray}	
A (\rho_0)&=&\frac{-1}{1+2\omega}2\Lambda
\left( (1-c+2c^2)+2c \left(c-{3\over 2}\right) Q_0\right)\label{f}\\[12pt]
B (\rho_0)&=&\frac{-1}{1+2\omega}
[(a+1)(e^{\beta Q_0}-1)-4\beta e^{\beta Q_0}]\rho_0\,.\label{g}
\end{eqnarray}
Equation~(\ref{e}) can be solved iteratively. At the first iteration:
\begin{equation}	
Q_1^{(1)}=-\frac{B}{A}
\end{equation}
and at the second and third iteration:
\begin{eqnarray}	
Q_1^{(2)}&=&\frac{1}{A}\square\left( \frac{B}{A}\right)-\frac{B}{A}\,,\\[12pt]
Q_1^{(3)}&=&-\frac{1}{A}\square\left( \frac{\square B/A}{A}\right) +
\frac{1}{A}\square \left( \frac{B}{A}\right)-\frac{B}{A}\,.
\end{eqnarray}
In order to have the correct dependence of the quantum potential on
the ensemble density, it is sufficient to set:
\begin{equation}	
A=k_1\sqrt{\rho_0}\,;\qquad B=k_2\rho_0 \label{h}
\end{equation}
where $k_1$ and $k_2$ are two constants.  This leads to the following
expressions for the quantum potential up to the third order of
iteration:
\begin{eqnarray}	
Q_1^{(1)}&=&-\frac{k_2}{k_1} \sqrt{\rho_0}\,, \\[12pt]
Q_1^{(2)}&=&\frac{k_2}{k_1^2} \frac{\square\sqrt{\rho_0}}{\sqrt{\rho_0}}-
\frac{k_2}{k_1}\sqrt{\rho_0}\,, \\[12pt]
Q_1^{(3)}&=&-\frac{k_2}{k_1^3} \frac{1}{\sqrt{\rho_0}}\square
\left( \frac{\square\sqrt{\rho_0}}{\sqrt{\rho_0}}\right)
+\frac{k_2}{k_1^2}\frac{\square\sqrt{\rho_0}}{\sqrt{\rho_0}}-
\frac{k_2}{k_1}\sqrt{\rho_0} \,.
\end{eqnarray}
If the ensemble density be not much great, and it be so smooth that
its higher derivatives be small, the result would be in agreement with
the desired relation $Q=\square\sqrt{\rho_0}/\sqrt{\rho_0}$ provided we
choose $k_2=k_1^2=k$.  Comparison of relations (\ref{f}),
(\ref{g}) and (\ref{h}) leads~to
\begin{eqnarray}	
a&=&2\omega k\,, \nonumber\\[10pt]
\beta &=&\frac{2\omega k+1}{4}\,, \\[12pt]
Q_0&=&\frac{1}{c(2c-3)} \left[ -\frac{2\omega k+1}{2\Lambda}k
\sqrt{\rho_0}-(2c^2-c+1)\right]\,.\label{a20}\nonumber
\end{eqnarray}
The space--time dependence of $\rho_0$ can be derived from the
relation (\ref{i}).

We see that the except $c$ and $\omega$, all other constants are
fixed.  The other equations of motion which are not used in the
perturbation procedure can be used to determine the space--time metric
and the Hamilton--Jacobi function.

We conclude this section by emphasizing on the fact that in our
present work, the quantum potential is a dynamical field. And, that
solving perturbatively the equations of motion, one gets the correct
dependence of quantum potential upon density plus some corrective
terms.

We stop here and shall investigate some physical results in the next section and after that continue our way to construct a theory for geometrization of quantum effects in terms of Weyl geometry.
\section{Some results of the idea}
In this section we shall first study some general solutions of  field
equations and then look for some specific cases like black holes and  big bang.
\subsection[Conformally flat solution]{Conformally flat solution\cite{qgg}\label{s5}}
Suppose we search for a solution which is conformally flat, and that
the conformal factor is near unity. Such a solution is of the form:
\begin{equation}      
g_{\mu\nu}=e^{2\Sigma}\eta_{\mu\nu};\qquad \Sigma\ll 1\,.
\end{equation}
As a result one can derive the following relations by equation (\ref{j}):
\begin{equation}
{\cal R}_{\mu\nu} = \eta_{\mu\nu}\square\Sigma+2\partial_\mu\partial_\nu\Sigma \Longrightarrow
{\cal G}_{\mu\nu}=2\partial_\mu\partial_\nu\Sigma -2\eta_{\mu\nu}\square\Sigma
\end{equation}
In order to solve for $\Sigma$ one can use the relation (\ref{k}),
and solve it iteratively as it is disscussed in the subsubsection (\ref{s14}). The result~is
\begin{eqnarray}            
{\cal R}^{(0)}&=&-\kappa {\cal T}\Longrightarrow
\Sigma^{(0)}=-\frac{\kappa}{6}\square^{-1}{\cal T}\,,\\[6pt]
{\cal R}^{(1)}&=&-\kappa {\cal T}+3\alpha\square\frac{\square\sqrt{|{\cal T}|}}{\sqrt{|{\cal T}|}}
\Longrightarrow 
\Sigma^{(1)} = -\frac{\kappa}{6}\square^{-1}{\cal T}+\frac{\alpha}{2}
\frac{\square\sqrt{|{\cal T}|}}{\sqrt{|{\cal T}|}}
\end{eqnarray}
and so on. Thus:
\begin{eqnarray}         
\Sigma&=&\underbrace{-\frac{\kappa}{6}\square^{-1}{\cal T}}_{\rm pure\ gravity}
\underbrace{+\frac{\alpha}{2}
\frac{\square\sqrt{|{\cal T}|}}{\sqrt{|{\cal T}|}}}_{\rm pure\ quantum}\nonumber\\[10pt]
&&{}+{\rm higher\ terms\ including\ gravity-quantum\ interactions}\,,
\end{eqnarray}
where $\square^{-1}$ represents the inverse of the dalambertian
operator. Note that the solution is in complete agreement with de
Broglie--Bohm theory.
\subsection[Conformally quantic solution]{Conformally quantic solution\cite{qgg}\label{s6}}
As a generalization of the solution found in the previous subsection,
suppose we set
\begin{equation}     
g_{\mu\nu}=e^{2\Sigma}\bar{g}_{\mu\nu}=(1+2\Sigma)\bar{g}_{\mu\nu};\qquad  \Sigma\ll 1\,.
\end{equation}
One can evaluate the following relations:
\begin{eqnarray}    
{\cal R}_{\nu\rho} &=& \bar{{\cal R}}_{\nu\rho}+\bar{g}_{\mu\nu}
\stackrel{-}{\square}\Sigma
+2\left( \bar{\nabla}_\nu\bar{\nabla}_\rho\Sigma+\bar{g}_{\nu\rho}
\bar{\nabla}_\alpha\Sigma
\bar{\nabla}^\alpha\Sigma-\bar{\nabla}_\nu\Sigma\bar{\nabla}_\rho
\Sigma\right)\,,\\[6pt]
{\cal G}_{\nu\rho} &=& \bar{{\cal G}}_{\nu\rho}-2\bar{g}_{\nu\rho}
\stackrel{-}{\square}\Sigma
+2\bar{\nabla}_\nu\bar{\nabla}_\rho\Sigma\,.
\end{eqnarray}
On using these relations in field equations (\ref{l}) one gets the following solution:
\begin{equation}      
\bar{{\cal G}}_{\mu\nu}=\kappa\bar{\cal T}_{\mu\nu};\ \ \ \ \Sigma=\frac{\alpha}{2}\Phi
\end{equation}
provided the energy--momentum tensor be conformally invariant, so under
this condition we have
\begin{equation}
g_{\mu\nu}^{\rm quantum+gravity}=(1+\alpha\Phi)g_{\mu\nu}^{\rm gravity}
\end{equation}
\subsection[Conformally highly quantic solution]{Conformally highly quantic solution\cite{qgg}\label{s7}}
Now we can generalize the result of the previous subsection. Suppose
in the overlined metric there is no quantum effect, so that $\bar{{\cal G}}_{\mu\nu}=\kappa\bar{{\cal T}}_{\mu\nu}$
and we know that the quantum effects could bring in via a conformal
transformation like $g_{\mu\nu}=e^{2\Sigma}\bar{g}_{\mu\nu}$.
Using field equations (\ref{l}) and the transformation properties
of the Einstein's equation one gets
\begin{equation}        
2\stackrel{-}{\square}\Sigma+2\bar{\nabla}_\alpha \Sigma
\bar{\nabla}^\alpha \Sigma =\alpha\stackrel{-}{\square}\Phi+
2\alpha\bar{\nabla}_\alpha \Phi\bar{\nabla}^\alpha
\Sigma -\frac{\alpha^2}{2}\bar{\nabla}_\alpha \Phi \bar{\nabla}^\alpha \Phi
\end{equation}
which has the solution
\begin{equation}        
\Sigma=\frac{\alpha}{2}\Phi\,.
\end{equation}
In the above solution it is assumed that the energy--momentum tensor
is either zero or conformally invariant.  So, under this condition, no
matter how large the quantum effects, are the general solution~is
\begin{equation}      
g_{\mu\nu}^{\rm quantum+gravity}=e^{\alpha\Phi}g_{\mu\nu}^{\rm gravity}\,.
\end{equation}
\subsection[Black holes]{Black holes\cite{qgg}\label{s8}}
Let us now use the above solutions to examine the quantum effects near the regions of
space--time where the gravitational effects of matter are large. Black
hole is the first examples we consider.

For a spherically symmetric black hole we have
\begin{equation}           
g_{\mu\nu}^{\rm gravity}=\text{diag}\left(
1-r_s/r,\frac{-1}{1-r_s/r},-r^2,-r^2\sin^2\theta\right)
\label{t}
\end{equation}
where $r_s$ is the Schwartzchid radius.
Using the fact that the Ricci scalar is zero for the above metric and the
transformation properties
of the Ricci scalar under conformal transformations, we have
\begin{equation}           
{\cal R} = 3\alpha e^{-\alpha\Phi}\left[ \stackrel{-}{\square}\Phi+\frac{\alpha}{2}
\bar{\nabla}_\mu \Phi\bar{\nabla}^\mu \Phi\right]
\end{equation}

\begin{figure}
\begin{center}
\epsffile{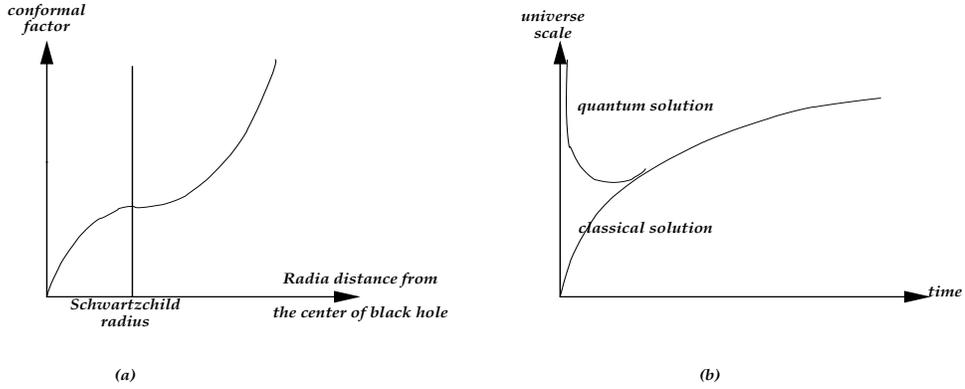}
\end{center}
\caption{(a) Conformal factor for a black hole, (b) The universe scale v.s. time.}
\label{fig1}
\end{figure}

The above equation is in fact a differential equation for the
conformal factor and can be solved for different regimes, giving the
following solution:
\begin{equation}      
g_{\mu\nu}^{\rm quantum+gravity}=g_{\mu\nu}^{\rm gravity}\times\left\{
\begin{array}{ll}
\exp\left(-\alpha r_s\over r^3\right)&r\rightarrow 0\,,\\[5pt]
{\rm Constant}&r\rightarrow r_s\,,\\[5pt]
\exp\left(r^2\over 3\alpha\right)&r\rightarrow\infty\,.
\end{array}\right.
\end{equation}
The conformal factor is plotted in Fig.~(\ref{fig1}-a). It can be seen
that the above conformal factor does not remove the metric singularity
at $r=0$.
\epsfxsize=5in \epsfysize=2in
\subsection[Initial singularity]{Initial singularity\cite{qgg}\label{s9}}
Big bang singularity is another place where  the gravitational effects are large. For an isotropic and homogeneous (FRW) universe we have
\begin{equation}
g_{\mu\nu}^{\rm gravity}=\text{diag}
\left(1,-\frac{a^2}{1-kr^2},-a^2r^2,-a^2r^2\sin^2\theta \right)
\end{equation}
As in the previous subsection we have
\begin{equation}
{\cal R} = e^{-\alpha\Phi}\left[\bar{{\cal R}}+ 3 \alpha\left(
\stackrel{-}{\square}\Phi+\frac{\alpha}{2} \bar{\nabla}_\mu \Phi
\bar{\nabla}^\mu \Phi\right)\right]
\end{equation}
As $t\rightarrow 0$ one can solve the above equations approximately
\begin{equation}     
g_{\mu\nu}^{\rm quantum+gravity}=g_{\mu\nu}^{\rm gravity}\exp\left({\alpha\over 2t^2}\right)
\end{equation}
so the universe scale is given by
\begin{equation}    
a(t)=a^{\rm classic}(t)\exp\left({\alpha\over 4t^2}\right)=\sqrt{t}\exp\left({\alpha\over 4t^2}\right)\,.
\end{equation}
As  can be seen easily, the curvature singularity at $t=0$ is removed
because as time goes to zero, the universe scale goes to infinity.
The behavior of the universe scale is ploted in Fig.~(\ref{fig1}-b).
\subsection[Production of inhomogenity]{Production of inhomogenity\cite{clu}\label{s10}}
At this end we shall examine the quantum effects on the motion of a fluid of matter. As a more practical example we take this fluid to be the cosmological matter fluid in an FRW model. We shall see that the quantum effects can produce some inhomogenities and thus produce cluster formation.  The hydrodynamics equation is given by:
\begin{equation}
\frac{\partial p}{\partial
x^\nu}g^{\mu\nu}+\frac{1}{\sqrt{-g}}\left ( \sqrt{-g}(p+\rho)U^\mu
U^\nu \right ) +\Gamma^\mu_{\nu\lambda}(p+\rho)U^\mu U^\nu= \rho \left ( g^{\mu\nu}-U^\mu U^\nu \right )
\frac{\partial\ln{\cal M}/m}{\partial x^\nu} \label{n}
\end{equation}
where we have introduced the quantum force in the right hand side
just as it is introduced in the equation of motion of a single
particle (see equation (\ref{m})). It must be noted that the
metric itself must be calculated from the corrected Einstein's
equations including the back--reaction terms. In
fact one must solve the above equation and the metric equation
simultaneously to obtain the metric and the density. We shall not
do in this way because solving those equations is difficult. We shall do in a similar
way. It is an iterative way and is based on the fundamental
assumption of this theory, that is equation (\ref{c}). As the
first order of iteration, we consider the space--time metric as
given by the classical Einstein's equations (Robertson--Walker
metric) and solve the above equation for the density, then using
the result obtained, calculate the quantum metric using equation
(\ref{c}). Then the new metric can be used to obtain the density
at the second order and so on.

In the comoving frame and with the assumption that the universe is
in the dust mode ($p=0$) with the flat Robertson--Walker metric,
we have from the equation (\ref{n}):
\begin{equation}
\frac{d\rho^{(1)}}{dt}+3H\rho^{(1)}=0 \label{o}
\end{equation}
\begin{equation}
\frac{\partial{\cal Q}^{(1)}}{\partial x^i}=0
\end{equation}
where $^{(1)}$ denotes the first order of iteration and
$H=\dot{a}/a$ is the Hubble's parameter. The solution of the above
two equations is:
\begin{equation}
\rho^{(1)}=\frac{{\cal X}^{(1)2}}{t^2};\ \ \ \ \ \ \ \ \ \ {\cal
Q}^{(1)}=\text{constant}
\end{equation}
where ${\cal X}^{(1)}$ should yet be determined. The constancy of
the quantum potential leads to:
\begin{equation}
{\cal
Q}^{(1)}= -\frac{\alpha}{2a^2} \frac{\nabla^2{\cal X}^{(1)}}{{\cal X}^{(1)}}
\end{equation}
so that:
\begin{equation}
\nabla^2 {\cal X}^{(1)}+\beta {\cal X}^{(1)}=0 \label{p}
\end{equation}
where:
\begin{equation}
\beta=\frac{2{\cal Q}^{(1)}a^2}{\alpha}
\end{equation}
This equation for ${\cal X}^{(1)}$ can simply be solved either in
the Cartesian coordinates or in the spherical ones. The solution
is:
\begin{equation}
{\cal X}^{(1)}=\sin\left ( \sqrt{\frac{\beta}{3}}x\right
)\sin\left ( \sqrt{\frac{\beta}{3}}y\right )\sin\left (
\sqrt{\frac{\beta}{3}}z\right ) \label{q}
\end{equation}
or:
\begin{equation}
{\cal X}^{(1)}=\sum_{l,m}\left ( a_{lm}j_l(\sqrt{\beta}r)+
b_{lm}n_l(\sqrt{\beta}r)\right ) Y_{lm}(\theta,\phi) \label{r}
\end{equation}
This is the first order approximation. At the second order, one
must use the equation (\ref{c}) to change the scale factor $a^2$
to $a^2(1+{\cal Q})$, and then from the relation (\ref{o}) we
have:
\begin{equation}
a^2(1+{\cal Q})=t^{2/3}{\cal X}^{(1)-4/3}
\end{equation}
So that:
\begin{equation}
{\cal Q}^{(2)}=-1+{\cal X}^{(1)-4/3}
\end{equation}
and then using this form of the quantum potential in the relation
(\ref{p}) or (\ref{q}) leads to the following approximation for
the density:
\begin{equation}
\rho^{(2)}=\frac{1}{t^2}\sin^2\left (
\sqrt{\frac{\gamma}{3}}x\right )\sin^2\left (
\sqrt{\frac{\gamma}{3}}y\right )\sin^2\left (
\sqrt{\frac{\gamma}{3}}z\right ) \label{s}
\end{equation}
where:
\begin{equation}
\gamma=\frac{2a^2}{\alpha}\left (-1+{\cal X}^{(1)-4/3}\right )
\end{equation}
and ${\cal X}^{(1)}$ is given by the relation (\ref{q}). This
procedure can be done order by order.

\begin{figure}
\begin{center}
\epsffile{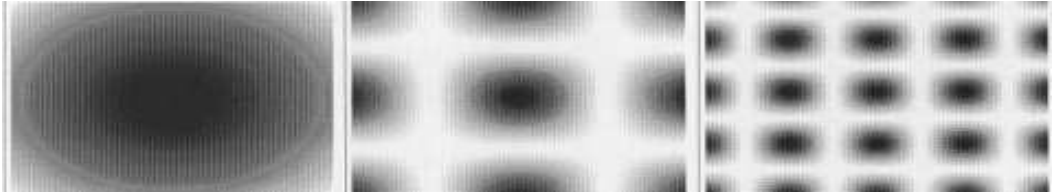}
\end{center}
\caption{Density at three stages of expansion for Cartesian solution}
\label{fig2}
\end{figure}

In the figures (\ref{fig2}) the density at three times are shown and the clustering
can be seen easily. These figures are plotted using the solution
in the Cartesian coordinates.

\begin{figure}
\begin{center}
\epsffile{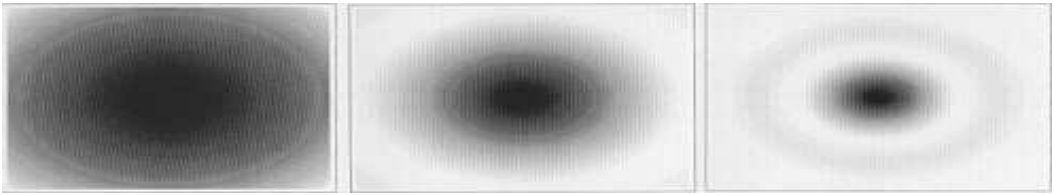}
\end{center}
\caption{Density at three stages of expansion for (00) mode}
\label{fig3}
\end{figure}

In figures (\ref{fig3}) the $(l,m)=(00)$ mode of equation (\ref{r}) is shown
at three time steps.

\begin{figure}
\begin{center}
\epsffile{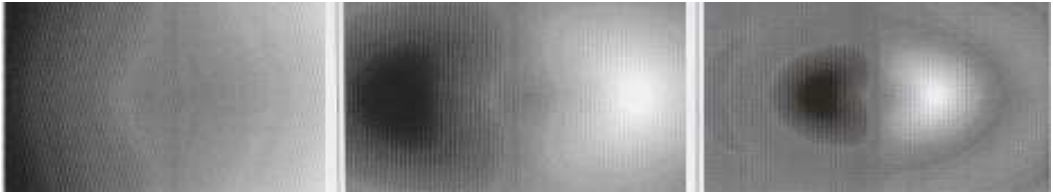}
\end{center}
\caption{Density at three stages of expansion for (11)$\oplus$(1-1) mode}
\label{fig4}
\end{figure}

In figures (\ref{fig4}) the $(11)\oplus (1-1)$ mode is shown at three time
steps.

\begin{figure}
\begin{center}
\epsffile{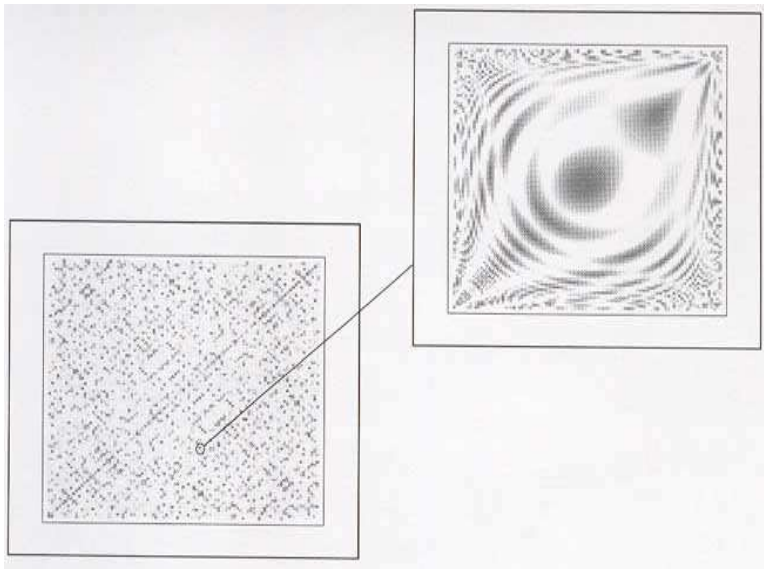}
\end{center}
\caption{The second order solution. Note the fine structure at the magnified portion.}
\label{fig5}
\end{figure}

The second order solution (\ref{s}) is shown in figure \ref{fig5},
where both large scale and small scale structures are shown.

It is important to note that the clustering can be seen in any of
these figures. In the last figure, however, one observes that at
the large scale the universe is homogeneous and isotropic, while
at the small scale these symmetries are broken.

\begin{figure}
\begin{center}
\epsffile{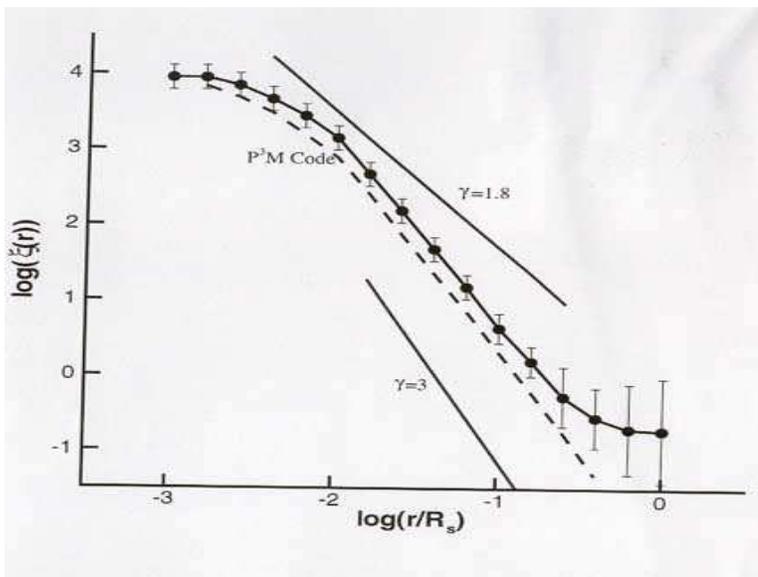}
\end{center}
\caption{Comparison with observation and $P^3M$ code.}
\label{fig6}
\end{figure}

At the end, in order to see whether our results are in agreement
with the observed clustering, the correlation function ($\xi(r)$)
is obtained from the  third order of iteration and is compared
with the cases $\xi=(r/r_0)^{-\gamma}$ with $\gamma=1.8$ and
$\gamma=3$ and with the standard result of a typical $P^3M$
code\cite{bor}. As it can be seen in figure (\ref{fig6}) our
results are in good agreement with the $P^3M$ code and with
observation.

Our claim here is not that this theory is
a good one for the cluster formation problem. But it is only
claimed that in the framework of causal quantum theory, the
quantum force \textit{may} be a cause for the cluster formation.
\epsfxsize=5.5in \epsfysize=1in
\epsfxsize=5.5in \epsfysize=1in
\epsfxsize=5.5in \epsfysize=1in
\epsfxsize=4in \epsfysize=3in
\epsfxsize=4in \epsfysize=3in
\subsection[Non--locality]{Non--locality\cite{non}\label{s11}}
Since the quantum potential in de-Broglie--Bohm theory has a highly non--local character\cite{bohm}, it is expected to see non--local behaviour of metric in this theory.
In order to illustrate how nonlocal effects can appear in quantum gravity
through quantum potential, suppose that matter distribution is localized
and has spherical symmetry. Then, one has:
\begin{equation}
\rho=\rho(t;r)
\end{equation}
\begin{equation}
\Omega=\Omega(t;r)
\end{equation}
Suppose, furthermore, that matter is at rest:
\begin{equation}
-\nabla_0S=E(t;r)\ \ as\ \ r\rightarrow \infty
\end{equation}
\begin{equation}
\nabla_iS=0;\ \ \ \ i=1,2,3
\end{equation}
One expects that at large $r$, where there is no matter, the background
metric would be of the Schwartzschield form. The validity of
this approaximation will be examined at the end. The equation of motion
(\ref{u}) relates $E$ and $\Omega$:
\begin{equation}
E=\frac{m\Omega}{\sqrt{1-r_s/r}}
\end{equation}
In order to calculate the conformal factor $\Omega$, one needs the specific
form of $\rho$. It must be a localized function at $r=0$. So we choose it as:
\begin{equation}
\rho(t;r)=A^2\exp[-2\beta(t)r^2]
\end{equation}
Using the relation (\ref{v}), the conformal factor can be simply calculated.
This leads to:
\[ \Omega^2=1+\alpha[\dot{\beta}^2r^4-\ddot{\beta}r^2+4\beta^2r] \]
from which we get:
\begin{equation}
\Omega^2\simeq \alpha\dot{\beta}^2r^4\ \ \ as\ \ r\rightarrow \infty
\end{equation}
Now it is a simple task to examine that the continuity equation (\ref{w})
is satisfied automatically as $r\rightarrow \infty$. This solution
is an acceptable one, only if the generalized Einstein's equations (\ref{x})
are satisfied. This is so if ${\cal T}^{(\Omega)}_{\mu\nu}\rightarrow 0$ as
$r\rightarrow\infty$. It can be shown that in the limit $r\rightarrow\infty$
we have:
\begin{equation}
\frac{\Box\Omega^2}{\Omega^2}=2(\ddot{\beta}/\dot{\beta})^2
+2\dot{\ddot{\beta}}/\dot{\beta}-20/r
\end{equation}
\begin{equation}
\frac{\nabla_0\nabla_0\Omega^2}{\Omega^2}=(\ddot{\beta}/\dot{\beta})^2
+\dot{\ddot{\beta}}/\dot{\beta}
\end{equation}
\begin{equation}
\frac{\nabla_1\nabla_1\Omega^2}{\Omega^2}=12/r^2
\end{equation}
\begin{equation}
\frac{\nabla_1\nabla_0\Omega^2}{\Omega^2}=(8\ddot{\beta}/r\dot{\beta})
\end{equation}
\begin{equation}
\left ( \frac{\nabla_0\Omega}{\Omega}\right )^2
=(\ddot{\beta}/\dot{\beta})^2
\end{equation}
\begin{equation}
\frac{\nabla_1\Omega\nabla_0\Omega}{\Omega^2}=(2\ddot{\beta}/r\dot{\beta})
\end{equation}
So provided that higher time derivatives of the scale factor of matter
density
($\beta$) are small with respect to its first time derivative,
that is:
\begin{equation}
\frac{\ddot{\beta}}{\dot{\beta}}\simeq0;
\frac{\dot{\ddot{\beta}}}{\dot{\beta}}\simeq 0\ \ and\ so\ on
\end{equation}
one has:
\begin{equation}
\lim_{r\rightarrow\infty}{\cal T}^{(\Omega)\nu}_{\mu}=0
\end{equation}
Also we have from (\ref{y}):
\begin{equation}
\lim_{r\rightarrow\infty}{\cal T}^{(m)\nu}_{\mu}=0
\end{equation}
So at large distances $g_{\mu\nu}$ satisfies Einstein's equations in vaccum,
${\cal G}_{\mu\nu}=0$. Therefore, the solution (\ref{t}) is acceptable. In this
way we find a solution to the quantum gravity equations at large distances.

Consequently, if the time variation of $\beta$ is small, the physical metric
$\tilde{g}_{\mu\nu}=\Omega^2g_{\mu\nu}$ is given by:
\begin{equation}
\lim_{r\rightarrow\infty}\tilde{g}_{\mu\nu}=\alpha\dot{\beta}^2r^4
g^{(Shwarzschield)}_{\mu\nu}
\label{z}
\end{equation}
An important points must be noted here. As it was shown, a change in
matter distribution (due to $\dot{\beta}$) instantaneosely alters the
physical metric. This is because of the appearance of $\dot{\beta}(t)$ in
equation(\ref{z}) and it comes from the quantum potential term.

We conclude that the specific form of the quantum potential leads to the appearance of
nonlocal effects in quantum gravity.
\section[Generalized equivalence principle]{Generalized equivalence principle\cite{weyl}\label{s12}}
After presenting some results of the idea of coding quantum effects in the conformal factor of the space--time metric, let us come back to our way of constructing a theory of it. One of the new points of the present approach for geometrization of quantum effects is the dual role of
geometry in physics. The gravitational effects determine the
causal structure of space--time as long as quantum effects
give its conformal structure. This does not mean that quantum
effects have nothing to do with the causal structure, they can act on
the causal structure through back--reaction terms appearing in the
metric field equations\cite{ss,st,conf}. We only mean that a
dominant role in the causal structure belongs to the gravitational
effects. The same is true for the conformal factor. The conformal
factor of the metric is a function of the quantum potential and  the mass of a relativistic particle is a field produced by quantum corrections to the classical mass.
We have shown  that the presence of the quantum
potential is equivalent to a conformal mapping of the metric. Thus
in conformally related frames we feel different quantum masses
and different curvatures. It is possible to
consider two specific frames. One of them contains the quantum
mass field (appeared in quantum Hamilton-Jacobi equation) and the
classical metric as while in the other the classical mass
(appeared in classical Hamilton-Jacobi equation) and the quantum
metric are appeared. In other frames both the space--time
metric and mass field have quantum properties. This
argument motivates us to state that \textit{different conformal
frames are identical pictures  of the gravitational and quantum
phenomena}. Considering the quantum force, the conformally
related frames aren't distinguishable. This is just what happens
when we consider gravity, different coordinate systems are
equivalent. Since the conformal transformation change the length
scale locally, we feel different quantum forces in different
conformal frames. This is similar to general relativity in which general
coordinate transformation changes the gravitational force at any
arbitrary point. Here it may be appropriate to state a basic
question. Does applying the above correspondence, between quantum and
gravitational forces, and between the conformal and general
coordinate transformations, means that the geometrization  of
quantum effects implies conformal invariance  just as
gravitational effects imply the general coordinate invariance?

To discuss this question, we  recall  what has been
considered earlier in the development of general
relativity. The general covariance principle leads to the
identification of gravitational effects of matter with the
geometry of the space--time. In general relativity the important
fact which supports this identification is the equivalence
principle. According to it, one can always remove the
gravitational field at some point by a suitable coordinate
transformation. Similarly, as we pointed out previously, according to our new
approach to Bohmian quantum gravity, at any point (or even
globally) the quantum effects of matter can be removed by a
suitable conformal transformation. Thus in that point(s) matter
behaves classically. In this way we can introduce a new
equivalence principle calling it the \textit{conformal equivalence
principle}, similar to the standard equivalence
principle. The latter interconnects gravity and general
covariance while the former has the same role about quantum
and conformal covariance. Both these principles state that
there is no preferred frame, either coordinate or conformal. Since Weyl geometry welcomes conformal invariance and since it
has additional degrees of freedom which can be identified with quantum effects, it
provides a unified geometric framework for
understanding the gravitational and quantum forces. In this way a pure geometric
interpretation of quantum behavior can be built.

Due to these results, we believe  that the de-Broglie--Bohm
theory must receive increasing attention in quantum gravity. This
theory has  some important features. One of them is that the
quantum effects appear independent of any preferred length scale (opposed to the standard quantum mechanics in which the Plank
length is a characteristic  length). This is one of the
intrinsic properties of this theory which results from a
special definition of the classical limit\cite{bohm}. Another
important aspect is that the quantum mass of the particle is a
field. This is needed for having conformal invariance, since mass
has a non--zero Weyl weight. Also, according to the geodesic equation (\ref{m}),
the appearance of quantum mass justifies Mach's principle\cite{mach}
which leads to the existence of interrelation between the global
properties of the universe (space--time structure, the large scale
structure of the universe,$\cdots$) and its local properties
(local curvature, motion in a local frame, etc.). In the
present theory, it can be easily seen that the space--time geometry is determined by the distribution of matter. A local
variation of matter field distribution changes the quantum
potential acting on the geometry. Thus the geometry is altered
globally (in conformity with Mach's principle). In this sense
our approach to the quantum gravity is highly non--local as it is
forced by the nature of the quantum potential\cite{non}. What we
call geometry is only the gravitational and quantum effects of
matter. Without matter the geometry would be meaningless. Moreover
in \cite{st,ss} we have shown that it is necessary to assume an
interaction term between the cosmological constant (large scale
structure) and the quantum potential (local phenomena). These
properties all justify Mach's principle. It is shown in
\cite{st,ss} that the gravitational constant is in fact a field
depending on the matter distribution through quantum potential.

All these arguments based on Bohmian quantum mechanics convince
us that Weyl geometry is a suitable framework for geometrization of
quantum mechanics.
\section[Formulation of the idea in terms of Weyl Geometry]{Formulation of the idea in terms of Weyl Geometry\cite{weyl}\label{s13}}
\subsection{Weyl--Dirac theory}
A straightforward generalization of the Einstein--Hilbert action to Weyl geometry leads
to a higher order theory\cite{we}. Dirac\cite{dir} introduced
a new action, called Weyl--Dirac action, by including a new field which is in
fact a gauge function. It helped him
to avoid higher order actions since fixing the gauge function led to
Einstein--Maxwell equations.

The Weyl--Dirac action is given by\cite{dir}:
\begin{equation}
{\cal A}=\int d^4x \sqrt{-g} \left ( F_{\mu\nu}F^{\mu\nu}-\beta^2\W{{\cal R}}+(\sigma +6)
\beta_{;\mu}\beta^{;\mu}+{\cal L}_{matter}\right )
\end{equation}
Here $F_{\mu\nu}$ is the curl of the Weyl four--vector $\phi_\mu$, $\sigma$ is an arbitrary constant, and
$\beta$ is a scalar field of weight $-1$. The ``$;$''
represents covariant derivative under general coordinate and
conformal transformations (Weyl covariant derivative) defined as:
\begin{equation}
X_{;\mu}=\W{\nabla}_\mu X-{\cal N}\phi_\mu X
\end{equation}
where ${\cal N}$ is the Weyl weight of $X$.
The equations of motion will then be:
\[
{\cal G}^{\mu\nu}=-\frac{8\pi}{\beta^2}({\cal T}^{\mu\nu}+M^{\mu\nu})+\frac{2}{\beta}(
g^{\mu\nu}\W{\nabla}^\alpha\W{\nabla}_\alpha\beta-\W{\nabla}^\mu\W{\nabla}^\nu\beta)
\]
\begin{equation}
 +\frac{1}{\beta^2}(4\nabla^\mu\beta\nabla^\nu\beta-g^{\mu\nu}\nabla^\alpha\beta \nabla_\alpha\beta)
+\frac{\sigma}{\beta^2}(\beta^{;\mu}\beta^{;\nu}-\frac{1}{2}g^{\mu\nu}\beta^{;\alpha} \beta_{;\alpha})
\end{equation}
\begin{equation}
\W{\nabla}_{\nu}F^{\mu\nu}=\frac{1}{2}\sigma(\beta^2\phi^\mu+\beta\nabla^\mu\beta) +4\pi J^\mu
\end{equation}
\begin{equation}
{\cal R}=-(\sigma+6)\frac{\W{\Box}\beta}{\beta}+\sigma\phi_\alpha\phi^\alpha
-\sigma \W{\nabla}^\alpha\phi_\alpha+\frac{\psi}{2\beta}
\end{equation}
where:
\begin{equation}
M^{\mu\nu}=\frac{1}{4\pi}\left ( \frac{1}{4}g^{\mu\nu}F^{\alpha\beta}F_{\alpha \beta} -F^\mu_\alpha F^{\nu\alpha}\right )
\end{equation}
and the energy--momentum tensor ${\cal T}^{\mu\nu}$, the current density vector $J^\mu$ and the scalar $\psi$ are defined as:
\begin{equation}
8\pi{\cal T}^{\mu\nu}=\frac{1}{\sqrt{-g}}\frac{\delta \sqrt{-g}{\cal L}_{matter}}{\delta g_{\mu\nu}}
\end{equation}
\begin{equation}
16\pi J^\mu=\frac{\delta {\cal L}_{matter}}{\delta \phi_\mu}
\end{equation}
\begin{equation}
\psi=\frac{\delta {\cal L}_{matter}}{\delta \beta}
\end{equation}
On the other hand the equations of motion of matter and the trace
of energy-momentum tensor can be obtained from the invariance of
the action under the coordinate and gauge transformations. One can
write them as:
\begin{equation}
\W{\nabla}_\nu{\cal T}^{\mu\nu}-{\cal T}\frac{\nabla^\mu\beta}{\beta}=J_\alpha\phi^{\alpha\mu}
- (\phi^\mu+\frac{\nabla^\mu\beta}{\beta})\W{\nabla}_\alpha J^\alpha
\end{equation}
\begin{equation}
16\pi{\cal T}-16\pi\W{\nabla}_\mu J^\mu-\beta\psi=0
\end{equation}
The first relation is only a geometrical identity (the Bianchi identity), and the
second shows the mutual dependence of the field equations.

It must be noted that in the Weyl--Dirac theory, the Weyl vector does not couple
to spinors, so $\phi_\mu$ cannot be interpretad as the electromagnetic potential\cite{lord}.
Here we use the Weyl vector not as the electromagnetic field but only as part
of the space--time geometry. The Weyl--Dirac formalism is adopted, and we shall
see that  the auxiliary field (gauge function) in Dirac's action represents the
quantum mass field. In addition both the gravitation fields ($g_{\mu\nu}$ and $\phi_\mu$)
and the quantum mass field determine the space--time geometry.
\subsection{Weyl--invariant quantum gravity}
In this section we shall construct a theory for Bohmian quantum
gravity which is conformally invariant in the framework of Weyl
geometry. To begin with, note that if our model should consider massive
particles, the mass must be a field. This is because mass has
non--zero Weyl weight. This is in agreement with Bohm's theory.
As we argued previousely a general Weyl invariant action is the
Weyl--Dirac action, whose equations of motion are derived in
the previous subsection. To simplify our model, we assume that the matter lagrangian
does not depends on the Weyl vector, so that $J_\mu=0$. The equations of motion are now:
\[
{\cal G}^{\mu\nu}=-\frac{8\pi}{\beta^2}({\cal T}^{\mu\nu}+M^{\mu\nu})+\frac{2}{\beta}(
g^{\mu\nu}\W{\nabla}^\alpha\W{\nabla}_\alpha\beta-\W{\nabla}^\mu\W{\nabla}^\nu\beta)
\]
\begin{equation}
 +\frac{1}{\beta^2}(4\nabla^\mu\beta\nabla^\nu\beta-g^{\mu\nu}\nabla^\alpha\beta
\nabla_\alpha\beta) +\frac{\sigma}{\beta^2}(\beta^{;\mu}\beta^{;\nu}-\frac{1}{2}g^{\mu\nu}\beta^{;\alpha}
\beta_{;\alpha})
\end{equation}
\begin{equation}
\W{\nabla}_{\nu}F^{\mu\nu}=\frac{1}{2}\sigma(\beta^2\phi^\mu+\beta\nabla^\mu\beta)
\label{aww}
\end{equation}
\begin{equation}
{\cal R}=-(\sigma+6)\frac{\W{\Box}\beta}{\beta}+\sigma\phi_\alpha\phi^\alpha
-\sigma \W{\nabla}^\alpha\phi_\alpha+\frac{\psi}{2\beta}
\label{xx}
\end{equation}
and the symmetry conditions are:
\begin{equation}
\W{\nabla}_\nu{\cal T}^{\mu\nu}-{\cal T}\frac{\nabla^\mu\beta}{\beta}=0
\label{zz}
\end{equation}
\begin{equation}
16\pi{\cal T}-\beta\psi=0
\label{yy}
\end{equation}
It must be noted that from equation (\ref{aww}) we have:
\begin{equation}
\W{\nabla}_\mu \left ( \beta^2\phi^\mu+\beta\nabla^\mu\beta\right )=0
\label{x13}
\end{equation}
so $\phi_\mu$ is not independent of $\beta$.

It is worthwhile to see whether or not  this model has anything to do with
the Bohmian quantum theory. We want to introduce the quantum mass field.
Now we    shall show that this field is proportional to the Dirac field. In
order to see this, two conditions should neccessary meet. Firstly, the correct
dependence of the Dirac field on the trace of energy--momentum tensor and, secondly
the correct appearance of the quantum force in the geodesic equation. Now note that using
equations (\ref{aww}),(\ref{xx}), and (\ref{yy}) we have:
\begin{equation}
\Box\beta+\frac{1}{6}\beta{\cal R}=\frac{4\pi}{3}\frac{{\cal T}}{\beta}+\sigma\beta
\phi_\alpha\phi^\alpha +2 ( \sigma-6)\phi^\gamma\nabla_\gamma \beta
+\frac{\sigma}{\beta} \nabla^\mu\beta\nabla_\mu\beta
\end{equation}
This equation can be solved iteratively.
Let we rewrite it as:
\begin{equation}
\beta^2=\frac{8\pi{\cal T}}{{\cal R}}-\frac{1}{{\cal R}/6-\sigma\phi_\alpha\phi^\alpha}
\beta\Box\beta +\cdots
\end{equation}
The first and the second order solutions of this equation are:
\begin{equation}
\beta^{2(1)}=\frac{8\pi{\cal T}}{{\cal R}}
\end{equation}
\begin{equation}
\beta^{2(2)}=\frac{8\pi{\cal T}}{{\cal R}}\left ( 1-\frac{1}{{\cal R}/6-\sigma\phi_\alpha\phi^\alpha}
\frac{\Box\sqrt{{\cal T}}}{\sqrt{{\cal T}}}+\cdots \right )
\label{ccc}
\end{equation}
To derive the geodesic equation we use the relation (\ref{zz}).
Assuming that matter consist of dust
with the energy--momentum tensor (\ref{aaa}) and multiplying equation (\ref{zz}) by $u_\mu$, we have:
\begin{equation}
u^\nu\W{\nabla}_\nu u^\mu=\frac{1}{\beta}(g^{\mu\nu} -u^\mu u^\nu ) \nabla_\nu\beta
\label{ddd}
\end{equation}
Comparison of equations (\ref{ccc}) and (\ref{ddd}) with equations (\ref{ee}) and (\ref{m})
shows that we have the correct equations of motions of Bohmian quantum theory, provided we identify:
\begin{equation}
\beta\longrightarrow {\cal M}
\end{equation}
\begin{equation}
\frac{8\pi{\cal T}}{{\cal R}}\longrightarrow m^2
\end{equation}
\begin{equation}
\frac{1}{\sigma\phi_\alpha\phi^\alpha-{\cal R}/6}\longrightarrow \alpha
\end{equation}
This shows that we have succesfully geometrized  Boh-mian quantum mechanics. The $\beta$ field is in fact the Bohmian quantum mass field, and the quantum coupling constant $\alpha$ which depends on $\hbar$ is also a field. In fact it is highly related to geometrical properties of the space--time through the above relation.

Since a gauge transformation can transform a general space--time dependent Dirac field to
a constant one, and vice-versa, it can be shown that quantum effects and the lenght
scale of the space--time are closely related. To see this suppose we are in a gauge
in which Dirac field is a constant. By applying a gauge transformation one can change
it to a general space--time dependent function.
\begin{equation}
\beta=\beta_0 \longrightarrow \beta=\beta(x)=\beta_0\exp (-\Xi(x))
\end{equation}
This gauge transformation is defined as:
\begin{equation}
\phi_\mu \longrightarrow \phi_\mu+\partial_\mu\Xi
\end{equation}
So, the gauge in which the quantum mass is constant (and thus the quantum force is zero)
and the gauge in which the quantum mass is space--time dependent are related to each other
via a scale change. In other words, $\phi_\mu$ in the two gauges differ by
$-\nabla_\mu(\beta/\beta_0)$. Since $\phi_\mu$ is a part of Weyl geometry, and Dirac field
represents the quantum mass, one concludes that the quantum effects are geometrized.
One can see this fact also by referring to the equation (\ref{aww}) which shows that
$\phi_\mu$ is not independent of $\beta$, so the Weyl vector is determined by quantum mass,
and thus this geometrical aspect of the manifold is related to the quantum effects.
 In this way, the physical meaning of auxiliary Dirac field is clarified, as while as
a suitable model for geometrization of quantum mechanics is introduced.
\subsection{Application to cosmology}
Most of physicists believe in a non--zero cosmological constant because of two important reasons.
It helps us to make the theoretical results to agree with observations. Morever some topics, like
large scale structure of the universe, dark matter, inflation, can be explored using it. On the
other hand from astronomical observations, especially gravitational lensing, cosmological constant
should be very small. ($|\Lambda|<10^{-54}/cm^2$) The fact that the cosmological constant is small
produces some difficulties. How explain theoretically this value of the cosmological constant? (This
question also applies to the gravitation coupling constant.) Morever the cosmological constant is a
measure of vaccum energy density. This includes some contribution from scalar fields, bare cosmological
constant, quantum effect, and so on. But observed cosmological constant is more smaller than (120 order of
magnitude less than) each one of the above contributions. This is the so--called cosmological constant
puzzle (see \cite{car} and its references). Till now many mechanisms are presented to solve the problem.

One way to solve the problem is to give dynamical characters to gravitational and cosmological constants
in such a way that they decrease as the universe expands.
Some works are done in \cite{hor} and \cite{bis}. In the former, a mechanism is presented using the
WDW equation, while the latter, focuses on the breaking the conformal invariance. Two scales, cosmological
and particle physics are introduced. And a dynamical conformal factor which relates them produces an effective
time dependent cosmological constant.

We also use the conformal invariance, but in the conformal invariant framework
of the present paper. Let's choose a spatially flat
Robertson--Walker metric:
\begin{equation}
ds^2=a^2(\eta)\left [ d\eta^2-dr^2-r^2d\Omega^2\right ]
\end{equation}
where $a(\eta)$ is the scale factor, and assuming the universe is filled of a dust,
the equations of motion of theory presented in the previous section now simplifies to:
\[
3\frac{\dot{a}^2}{a^4}-\frac{8\pi\rho}{\beta^2}+\frac{6}{\beta}\left ( \frac{\dot{a}}{a}-\phi\right ) \frac{\dot{\beta}}{a^2}+
\frac{3}{\beta^2}\frac{\dot{\beta}^2}{a^2}
\]
\begin{equation}
+\frac{\sigma}{2\beta^2}\frac{(\dot{\beta}+\phi\beta)^2}{a^2}=0
\end{equation}
\begin{equation}
\dot{\beta}+\beta\phi=0
\end{equation}
\[
-6\frac{\ddot{a}}{a^3}-(\sigma+6)\left ( \frac{1}{\beta}\frac{d}{d\eta}\left (\frac{\dot{\beta}}{a^2}\right ) +\frac{\dot{\beta}}{\beta a^2}
\left (4\frac{\dot{a}}{a}-10\phi\right ) \right )
\]
\begin{equation}
+\sigma \frac{\phi^2}{a^2}-\sigma \frac{d}{d\eta}\left (\frac{\phi}{a^2}
\right ) -\sigma \frac{\phi}{a^2}
\left ( 4\frac{\dot{a}}{a}-10\phi\right )+\frac{\psi}{2\beta}=0
\end{equation}
where a dot over any quantity represents derivation with respect to time and we have chosen the gauge
\begin{equation}
\phi_\mu=(\phi,0,0,0)
\end{equation}
And the symmetry conditions are:
\begin{equation}
\dot{\rho}+3\rho\left ( \frac{\dot{a}}{a}-\phi\right )-\rho\frac{\dot{\beta}}{\beta}=0
\end{equation}
\begin{equation}
16\phi\rho-\beta\psi=0
\end{equation}
Introducing the cosmological time as $dt=ad\eta$ and simplifying the relations, we finally have:
\begin{equation}
\rho a^3\beta^2=constant
\end{equation}
\begin{equation}
3\frac{a'^2}{a^2}-\Lambda_{eff}-8\pi G_{eff}\rho=0
\end{equation}
\begin{equation}
3\frac{a''}{a}+3\frac{a'^2}{a^2}+30\frac{\beta'^2}{\beta^2}+9\frac{a'}{a}\frac{\beta'}{\beta}+3\frac{\beta''}{\beta}-4\pi G_{eff}\rho=0
\end{equation}
where a $'$ over any quantity represents derivation with respect to the cosmological time and we have deffined:
\begin{equation}
\Lambda_{eff}=-9\frac{\beta'^2}{\beta^2}-6\frac{a'}{a}\frac{\beta'}{\beta}
\end{equation}
\begin{equation}
G_{eff}=\frac{1}{\beta^2}=\frac{1}{{\cal M}^2}
\end{equation}
The above equations can simply solved resulting in:
\begin{equation}
H\sim t^{-1}
\end{equation}
\begin{equation}
\Lambda_{eff}\sim t^{-2}
\end{equation}
\begin{equation}
G_{eff}\sim t^{-4/19}
\end{equation}
where $H$ is the Hubble constant. As the universe expands these quantities
decrease in agreement with the above disscusion. These constants
have a small value at the current epoch as the observation suggests.
It can be noticed that these time dependences are through the quantum mass field ($\beta$ or ${\cal M}$). The quantum mass field changes with time as $t^{2/19}$.
\section{Extension of the results\label{s20}}
In this section we shall look for the possibility of extending the theory in two lines. Manipulating many--particle systems and inclusion of spin.
\subsection{Many--particle systems}
Till now our disscusion was about geometrization of quantum effects of single particle systems. What happens when one deals with a system consisting of more than one particle? First the quantum potential for many--particle systems in the non--relativistic case is given by\cite{bohm}:
\begin{equation}
Q=-\frac{\hbar^2}{2m}\frac{1}{|\psi|}\sum_{i=1}^N\nabla_i^2|\psi|
\end{equation}
Its generalization to the case of particles moving in an arbitrary space--time is clearly:
\begin{equation}
Q=\frac{\hbar^2}{m^2}\frac{1}{|\psi|}\sum_{i=1}^N\Box_i^2|\psi|
\end{equation}
A difficulty arrises here. This quantum potential is defined in the configuration space, i.e. it is a function of $x_1$, $x_2$, $\cdots$ $x_N$. Simply putting the exponential of the quantum potential as the conformal factor of the space--time metric should not work because this makes the metric to live in the configuration space, a completely meaningless thing. To solve the problem we can generalize the idea in this way. \textit{Different particles of the many--particle system does not experience the same geometry. The $i$th particle sees the metric:}
\begin{equation}
g_{\mu\nu}(x_i)=\left . \exp\left ( Q\right )\overline{g}_{\mu\nu}\right |_{x_j=r_j\ \text{for}\ j\neq i}
\end{equation}
\textit{where $r_j$ denotes the Bohmian trajectories.} In this way one can see that there is no problem in geometrization of quantum effects for many--particle systems.
\subsection{Spin}
There are at least two approaches to include spin in de-Broglie--Bohm theory. First one can see the spin as the $rotation$ of the particle around itself\footnote{This can lead to superluminal angular velocities.}. This approach can easily used here. Only one needs to add the rotational degrees of freedom for the motion of the particle. This is surely straightforward.

The other approach looks at the spin (just like the Copenhagen quantum mechanics) as an internal degree of freedom. So the spin is coded in the wave function. There is also no difficulty in adopting this approach.  To explain this, first note that using the Wigner formulation of any spin wave equation\cite{wig,lord}, one has a Dirac--like wave equation for any spin:
\begin{equation}
\left ( \gamma^\mu\partial_\mu\right )^\beta_{\alpha_1}\psi_{\beta\alpha_2\cdots \alpha_{2s}}=\frac{im}{\hbar}\psi_{\alpha_1\cdots \alpha_{2s}}
\end{equation}
where the Dirac spinor of rank $2s$ is completely symmetric in its spinor indices. For zero mass we also has the condition $\gamma^{5\beta}_{\alpha_1}\psi_{\beta\alpha_2\cdots \alpha_{2s}}=\psi_{\alpha_1\cdots \alpha_{2s}}$.
The velocity field compatible with this wave equation is the Dirac current field $j^\mu$.
Now one must add the continuity equation which can be seen as an equation for defining the ensemble density $\rho$, and define the quantum potential in terms of it as usual.
So the space--time metric is $g_{\mu\nu}=\exp(Q)\overline{g}_{\mu\nu}$.
\section{Conclusion}
We have used an approach which is different from other existing
ones, which try to combine the gravitational and quantal effects.  By
the investigation of the quantum effects of matter in the framework of
Bohmian
mechanics, we have shown that the motion of a particle with quantum
effects is equivalent to its motion in a curved space-time.  We have
investigated the coupling of purely gravitational effects and purely
quantal
effects of the particle, by considering a general background space-time
metric. The use of the de-Broglie--Bohm quantum theory of motion,
instead of the standard Copenhagen quantum mechanics, has
at least three advantages. First, that the inherent problems of the
standard quantum mechanics are not present. Second, that the conceptual
problems of the standard quantum gravity like the meaning of the
wave-function, are circled. Finally, the
equivalence of quantum effects of matter and a curved space-time,
which is our most important result, is
achieved through this point of view. This leads directly to the
minisuperspace of conformal degree of freedom, in which, the conformal
factor has now a clear physical meaning. Other problems like, time-
dependence
which is necessary to understand the evolution of the universe,
consideration of the back--reaction effects, and so on, are also handled
in the present theory.
An important property of this theory is that the conservation laws
are the same as those of the classical theory.
As it is shown, if one applies this theory to the
case of quantum cosmology, it is possible to solve the equations of
motion nonperturbatively (i.e. exactly). One sees that there is no
singularity at small times
(provided one assumes that the quantum coupling constant of radiation is
negative). It is smeared out by the quantum effects.
Two remarks are in order here. First, in principle, the above model
can be applied to the non flat Friedmann universes
and similar results would emerge. Second, in the present theory, the
quantum
effects are only those of matter. The effect of these quantal behaviours
on the background space-time is achieved via the modified Einstein
equations, i.e. via the back--reaction effects. This means that if one
removes
the matter, the quantum effects of the background geometry would
disappear. In order to generalize the present theory to a fully
theory of quantum gravity, the quantum effects of gravity must be
included.

As explained before the keystone of Bohm's theory is the \textit{quantum potential}. Any particle is acted upon by a quantum force
derived from the quantum potential. The quantum potential is itself
resulted from some self-field of the particle, the wave function.
Since the quantum potential is related only to the norm of the wave
function and because of Born's postulate asserting that the ensemble
density of the particle under consideration is given by the square of
the norm of the wave function, the quantum potential is obtained
resulted from the ensemble density. The non understandable point of
Bohm's theory is just this. How does a particle know about its
hypothetical ensemble? When the hypothetical ensemble is a real one,
i.e.~when there are is large number of similar particles just like the
particle under consideration, quantum potential can be understood. It
is a kind of interaction between the particles in the real
ensemble. But when one deals with only one particle the quantum
potential is interaction with the other hypothetical particles!!!

On the other hand, quantum potential is highly related to the
conformal degree of freedom of the space--time metric. In fact, the
presence of the quantum force is just like having a curved
space--time which is conformally flat and the conformal factor is
expressed in terms of the quantum potential. In this way one sees that
quantum effects are in fact geometric effects. Geometrization of
the quantum theory can be done successfully, but still, there is the
problem of the ensemble noted above.

Here we have shown that if one tries to geometrize the
quantum effects in a purely metric way, the ensemble problem would be
overcome. In addition it provides the framework for bringing in the
purely quantum gravity effects.

A point
 about the geodesic equation must be noted here. In the
background metric, this equation resembles the geodesic equation in
Brans--Dicke theory. Consideration of the matter quantum effects, leads
to the physical metric in which a particle moves on the geodesic of
Branse--Dicke theory written in Einstein gauge. This point supports
the suggestion that the discussion of quantum gravity requires a
scalar--tensor theory. Previously this was suggested when discussing
 Bohmian quantum gravity \cite{cov}.

Next it is shown that it is possible to write a scalar--tensor theory
which automatically leads to the correct equations of motion. This has
the
advantage that the conformal factor would be fixed by the equations of
motion, and not by introducing a lagrangian multiplier by hand.
We see that  the matter distribution determines the local
curvature
of the space--time (in confirmity with Mach's principle). Furthermore,
from the matter equation of motion one can see that the cosmological
constant
(a large scale structure constant) and the quantum potential\footnote{In
Bohmian quantum mechanics, observable effects of quantum potential
may appear at both large and small scales, depending on the shape of
the ensemble density\cite{bohm}.} are coupled together.
This is another manifestation of the Mach's principle.

Also  we have constructed a scalar--tensor theory with two
scalar fields, for which the equations of motion lead to the correct
form of quantum potential. We have not only shown that
quantum effects are geometrical in nature, but also \textit{derive} the
form of quantum potential. This specific form for quantum potential is
a result of the equations of motion.

It must be noted that in this theory both the scalar fields interact
with the cosmological constant.  Hence, the presence of the
cosmological constant (even very small) is essential in order the
theory works.  Note that the interaction between the cosmological
constant and quantum potential represents a connection between the large
and small scale structures.

We presented a toy model, and investigated its solutions. Also it is shown
that the initial singularity is removed by quantum effects.

Finally we saw that one can
formulate a \textit{generalized equivalence principle} which states
that gravitation can be removed locally via an appropriate
coordinate transformation, while quantum force can be removed
either locally or globally via an appropriate scale
transformation. So the natural framework of quantum and gravity is
Weyl geometry. The most
simplest Weyl invariant action functional is written out. It
surprisingly leads to the correct Bohm's equations of motion. When
it applied to cosmology it leads to time decreasing cosmological and
gravitational constants.
A phenomena which is good for describing their small values.

\end{document}